\documentclass[%
 reprint,
 amsmath,amssymb,
 aps,
]{revtex4-2}

\usepackage{graphicx}% Include figure files
\usepackage{dcolumn}% Align table columns on decimal point
\usepackage{bm}% bold math
\usepackage[utf8x]{inputenc}

\newcommand{\D}[1]{\ensuremath{\text{d}#1}}

\newenvironment{equationno*}
{\begin{linenomath*}
\begin{equation*}}
{\end{equation*}
\end{linenomath*}}

\begin{document}

\title{An Analytically Solvable Model of Firing Rate Heterogeneity in Balanced State Networks}% Force line breaks with \\

\author{Alexander Schmidt\textsuperscript{1-3}}
\author{Peter Hiemeyer\textsuperscript{1}}
%\author{Shy Shoham}\textsuperscript{4}
\author{Fred Wolf\textsuperscript{1-7}}
 \email{fred.wolf@ds.mpg.de}
\affiliation{$^1$Max Planck Institute for Dynamics and Self-Organization, Göttingen, Germany}
\affiliation{$^2$Göttingen Campus Institute for Dynamics of Biological Networks, University of Göttingen, Göttingen, Germany}
\affiliation{$^3$Max Planck Institute for Multidisciplinary Sciences, Göttingen, Germany}
\affiliation{$^4$Institute for the Dynamics of Complex Systems, University of Göttingen, Göttingen, Germany}
\affiliation{$^5$Center for Biostructural Imaging of Neurodegeneration, Göttingen, Germany}
\affiliation{$^6$Bernstein Center for Computational Neuroscience Göttingen, Göttingen, Germany} 
\affiliation{$^7$Cluster of Excellence "Multiscale Bioimaging: from Molecular Machines to Networks of Excitable Cells" (MBExC), University of Göttingen, Göttingen, Germany}

% Primary Equal Contribution Note
%\Yinyang These authors contributed equally to this work.
\maketitle

%\begin{widetext}
%This is some test of an abstract
%\end{widetext}
%\section*{Abstract}
    \textbf{
        Distributions of neuronal activity within cortical circuits are often found to display highly skewed shapes with many neurons emitting action potentials at low or vanishing rates, while some are active at high rates. Theoretical studies were able to reproduce such distributions, but come with a lack of mathematical tractability, preventing a deeper understanding of the impact of model parameters. In this study, using the Gauss-Rice neuron model, we present a balanced-state cortical circuit model for which the firing rate distribution can be exactly calculated. It offers selfconsistent solutions to recurrent neuronal networks and allows for the combination of multiple neuronal populations, with single or multiple synaptic receptors (e.g. AMPA and NMDA in excitatory populations), paving the way for a deeper understanding of how firing rate distributions are impacted by single neuron or synaptic properties.
    }

\section{Introduction}
    
    Spikes are the principal currency of information flow in the cerebral cortex, but cortical neurons differ strongly in the amount of this currency they are handing out per time unit. In general, firing rates in \textit{bona fide} uniform populations of cortical principal neurons vary widely and often exhibit highly skewed distributions both in spontaneous and in driven states \cite{hirase_firing_2001,battaglia_firing_2005,shafi_variability_2007,hromadka_sparse_2008,oconnor_neural_2010,peyrache_spatiotemporal_2012,mizuseki_preconfigured_2013,Busche2015,shoham_how_2006,buzsaki_log-dynamic_2014}. It is well understood that broad and skewed firing rate distributions are a generic and robust property in models of recurrently connected networks of cortical neurons \cite{van_vreeswijk_chaos_1996,van_vreeswijk_chaotic_1998,roxin_distribution_2011,monteforte_dynamical_2010,monteforte_dynamic_2012,renart_asynchronous_2010,amit_dynamics_1997,brunel_dynamics_2000}. In particular, in balanced state networks broad and skewed firing rate distributions even emerge in situations of uniform external drive. This results because most inputs to individual neurons originate from other neurons within the cortical circuit, making the total input subject to heterogeneities in the composition of cortical presynaptic cells and the total number of received synapses \cite{van_vreeswijk_chaos_1996,van_vreeswijk_chaotic_1998}. Together with the approximate balancing of excitation by feedback inhibition, this generically leads to the emergence of broad and skewed firing rate distributions. 
    
    The shape and range of the emerging firing rate distribution, in general, is determined by a set of self-consistency requirements \cite{van_vreeswijk_chaos_1996,van_vreeswijk_chaotic_1998}. In essence, inputs drawn from the correct firing rate distribution must drive the postsynaptic cells in the network to fire at rates that agree with the assumed firing rate distribution. For balanced state networks on random graphs, these conditions can be expressed mathematically by a few self-consistency equations that become exact in the large system limit. Prior work indicates that these equations are sensitive to a variety of network parameters, from synaptic efficiencies and connection probabilities to single neuron f-I-curves and the dynamics of synaptic inputs \cite{van_vreeswijk_chaos_1996,van_vreeswijk_chaotic_1998,roxin_distribution_2011,monteforte_dynamical_2010,monteforte_dynamic_2012,renart_asynchronous_2010,amit_dynamics_1997,brunel_dynamics_2000}. These seminal studies conclusively demonstrated that the emergence of realistic firing rate distributions is a robust feature of such networks. Due to the complexity of the dependencies on biological detail and a lack of mathematical tractability of the self-consistency equations, the impact of synaptic and network parameters on the shape and range of firing rate distributions is not sufficiently understood. 

    Utilizing the superior mathematical tractability of the Gauss-Rice neuron model, we here introduce a balanced-state cortical circuit model for which the firing rate distribution can be exactly calculated. We find that the self-consistency equations determining the shape and position of the firing rate distribution take the form of relatively simple transcendental equations that can be easily solved. The emergent firing rate distribution is determined by a small number of effective parameters that capture the dependence of the rate distribution on the cellular and synaptic properties in the network. In particular, we treat the case of a local circuit composed of excitatory and inhibitory neurons in which excitatory synaptic currents exhibit different decay times reflecting for instance a mixture of AMPA and NMDA receptors at individual synapses.
    
\section{Methods}\label{sec: methods}

    \subsection{Network topology and setup}
% basic ER-setup
    We consider a network defined by a directed Erd\H{o}s-R\`{e}nyi graph with first order connectivity $p=K/N$, where $K$ is the average number of connections per neuron and $N$ the total number of neurons in the network. The limit $1 \ll K \ll N$, with $N,K \rightarrow \infty$, allows for a mean-field approach when modeling network dynamics.
    
% connectivity matrix
    The connectivity is  described by an $N\times N$-matrix with entries $A^{ij}_{kl}$ that encode the connections of neuron $j$ of population $l$ to neuron $i$ of population $k$. The population indices $k$ and $l$ point to regions of the connectivity matrix, that refer to the respecive population and that follow statistics based on their identity. The indices $i$ and $j$ run only over their respective populations. Differences in absolute numbers of synaptic connections between the populations can be captured by the factor $\kappa_l$. $A^{ij}_{kl}=1$ if there is a connection and $0$, otherwise. The mean outward connectivity of population $l$ is obtained as
    \begin{equation*}
      [A_{kl}^{ij}]_{ijk}=\sum_{k=EI}\frac{1}{N_k}{\sum_{i=1}^{N_k}\frac{1}{N_l}\sum_{j=1}^{N_l}{A^{ij}_{kl}}}\;,
    \end{equation*}
    which is an average over all presynaptic neurons of population $l$ and all postsynaptic neurons in the network. It yields $[A^{ij}_{kl}]_{ijk} = \kappa_l\frac{K}{N}$.
    
    \begin{figure}
      \includegraphics[width=\linewidth]{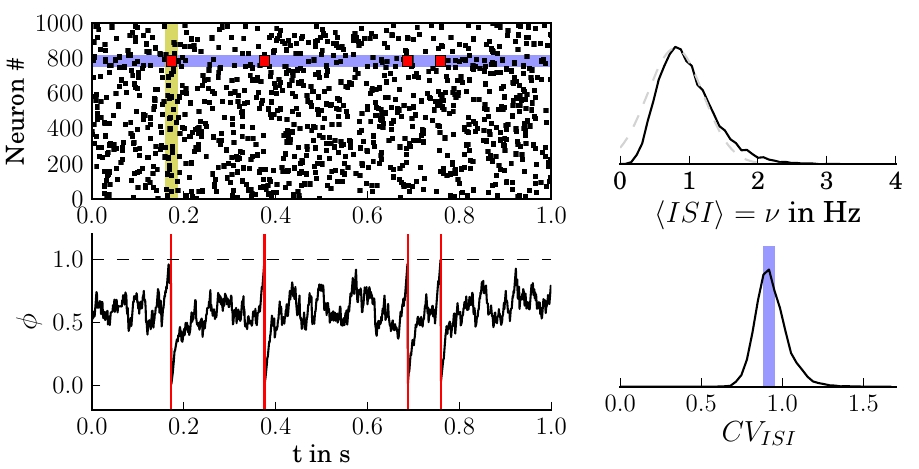}
      \caption[Asynchronous, irregular spiking activity in the Balanced State]{Asynchronous, irregular spiking activity in the Balanced State; top left: spike raster of a network. Irregular spiking activity of one neuron over time is highlighted by the blue bar while the yellow bar highlights the asynchronous activity over the network at a random timepoint; bottom left: phase trace of the highlighted neuron; top right: the firing rate distribution has a characteristic shape with a peak at low rates and some skewness (grey dashed gaussian as reference); bottom right: Irregular firing is indicated by a CV value distribution peaked close to one}
    \label{fig: balancedState}
    \end{figure}
    
% synaptic weight scaling
    A stationary state with temporal fluctuations independent of network size is obtained by scaling the synaptic weights according to \cite{softky_highly_1993,van_vreeswijk_chaos_1996,van_vreeswijk_chaotic_1998}:
    \begin{equation}
    	\tilde{J}_{kl} = \pm \frac{J_{kl}}{\sqrt{K}}\,,\qquad J_{kl}>0\,. \label{eq: syn_weight}
    \end{equation}
    
% difference in inhib / exc. neurons
    Interactions between neurons are mediated by electrical signals (action potentials or \textit{spikes}) inducing post-synaptic currents (PSCs), that can be of either inhibitory (negative synaptic weight in eq.~\ref{eq: syn_weight}, $l = I$) or excitatory (positive weight, $l = E$) nature. According to Dale's Principle, our model neurons can only have one kind of impact on its postsynaptic neurons, allowing for a clear distinction of two populations in the network: an excitatory (E) and an inhibitory (I) one.
    
    % PSCs
    We describe PSCs by a normalized kernel of shape
    \begin{align}
	f_l(t-t^{(p)}) &= \sum_m \frac{r_{l_m}}{\mathcal{N}_{l_m}}\exp{\left(-\frac{t-t^{(p)}}{\tau_{I_{l_m}}}\right)}\Theta(t-t^{(p)}) \label{eq: kernel}\\
	\sum_m r_{l_m} &= 1 \,,\quad \mathcal{N}_{l_m} \in [1,\tau_{I_{l_m}} ] \nonumber
    \end{align}
    where $\Theta$ is the Heavyside-function and $t^{(p)}$ the time of a spike. Our model allows for multiple types of post-synaptic receptors, including mixtures with different temporal characteristics: $\tau_{I_{l_m}}$ is the $m$th synaptic timeconstant of population $l$, e.g. inhibitory (GABAergic, few ms decay time, \cite{otis_modulation_1992,sigel_structure_2012}) or fast (AMPA, few ms, \cite{attwell_neuroenergetics_2005}) and slow (NMDA, several $100\,$ms, \cite{attwell_neuroenergetics_2005,traynelis_glutamate_2010}) excitatory ones. The mixture parameters $r_{l_m}$ specify the fraction of each type of receptor per synapse and the normalization constant $\mathcal{N}_{l_m}$ specifies, whether the kernel is normalized with respect to the total deposited charge ($=\tau_{I_{l_m}}$), or to its peak current  ($=1$). In the limit $\tau_I \rightarrow 0$ one would obtain $\delta$-pulses, effectively resulting in an Ornstein-Uhlenbeck process membrane potential process.
    
    For simplicity, we first assume \textit{pure} kernels first, with a single post-synaptic receptor type:
    \begin{equation}
	    f_l(t-t^{(p)}) = \frac{1}{\mathcal{N}_l}\exp{\left(-\frac{t-t^{(p)}}{\tau_{I_{l}}}\right)}\Theta(t-t^{(p)})\,, \label{eq: kernel_single}
    \end{equation}
    and later treat the case of mixed receptors in Sec.~\ref{sec: methods_mixed_kernel}
    
    We assume uniform synaptic weights and kernels of the PSCs for all neurons within a population and finally drive the network by a population-specific, external constant excitatory current $I_k^{\text{ext}}$.
    
    As we will show below, this setup allows calculating of quenched and dynamic fluctuations of inputs across the network (at any given time) once the compound spike train statistics are determined (Fig.~\ref{fig: way_to_current}(c)). 
    
    \subsection{Input current statistics}
        
        \subsubsection{Compound spike train statistics}\label{sec: compound_spike_train}
    \begin{figure}
      \centering
      \includegraphics[width=\linewidth]{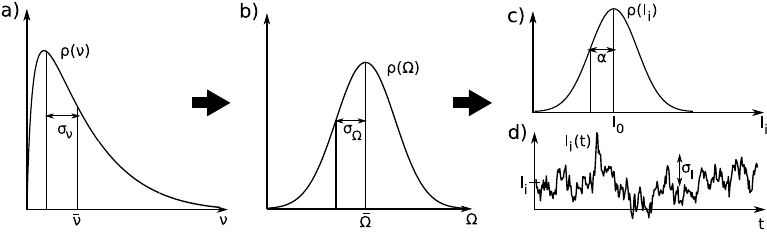}
      \caption{Spike train- and input current statistics are derived from the firing rate distribution; (a) probability density function (pdf) of firing rates across neurons $\rho(\nu)$; (b) pdf of compound spike rate $\Omega$; (c) input current distribution (mean and quenched variance marked); (d) temporal fluctuations of input current governed by the shape of the PSC-kernel}
      \label{fig: way_to_current}
    \end{figure}
    
    The temporal sequence of spikes from a single neuron constitutes a point process \cite{softky_highly_1993,compte_temporally_2003,kerr_imaging_2005}, termed \textit{spike train}. The entirety of incoming spikes to neuron $i$ is the \textit{compound spike train} with rate
    \begin{equation}
        \Omega_i(t) = \sum_{j\in \text{pre(i)}} \nu_j(t)\,, \label{eq: methods_compound_spike_train}
    \end{equation}
    where pre(i) is the set of presynaptic neurons to neuron $i$ ($A_{kl}^{ij} = 1$). Individual firing rates $\nu_j$ are sampled from an underlying, yet unknown firing rate distribution (Fig.~\ref{fig: way_to_current}(a)), which is to be selfconsistently derived. Due to $N\rightarrow \infty$ and the sparse connectivity ($N\gg K$) presynaptic neuronal activity is uncorrelated \cite{renart_asynchronous_2010} and the compound spike train can be approximated as a Poisson-like point process, even if individual cells' activity deviates from this simple statistics. Large $K$ by application of the central limit theorem implies the total recurrent synaptic current to be Gaussian. Accordingly, the \textit{Compound spike train statistics} (Fig.~\ref{fig: way_to_current}(b)) are sufficiently described by their first and second moment.
        
    In the stationary case the spike train of each presynaptic neuron pre(i) contributes on average $\nu$ APs per second to the Poissonian compound spike rate $\Omega_i$ (eq.\ref{eq: methods_compound_spike_train}) with the network average:
    \begin{align}
        \Omega &= [\Omega_i]_i = \sum_{l=E,I} [\Omega_l^i]_i \nonumber \\
        &=\sum_{l=E,I}\left[\sum_{j=1}^{N_l} A_{kl}^{ij}\nu_{l}^{j}\right]_i
        =\sum_{l=E,I} N_l \left[A_{kl}^{ij}\right]_{ij}\left[\nu_l^{j}\right]_{j}  \nonumber\\
        & = \sum_{l=E,I} N_l \kappa_l \frac{K}{N_l}\bar{\nu}_l = K (\kappa_E \bar{\nu}_E + \kappa_I \bar{\nu}_I)\,, \label{eq: omega_mean}
    \end{align}
    where $[\cdot]_i$ denotes the network average over postsynaptic neurons $i$ and we used the statistical independence of presynaptic firing rates $\nu_l^j$ and the matrix entries $A_{kl}^{ij}$. The second moment $[\Omega_{i}^{2}]_i$ and thus the resulting variance over the network $\text{Var}\left(\Omega\right) = [\Omega_{i}^{2}]_i-[\Omega_{i}]_i^{2}$ evaluates to (see App.\ref{sec: appendix_compound_spike_train_variance}):
    \begin{align}
        \text{Var}\left(\Omega\right) &= K (\kappa_E q_E + \kappa_I q_I) \,, \label{eq: omega_var}
    \end{align}
    where $q_l$ denotes the second moment of the firing rate distribution of population $l$. This concludes the calculations of the statistics of the compound spike train statistics, Fig.~\ref{fig: way_to_current}(b)
    
    %In the large network limit the statistics are set by the first and second moment. 
      
    \subsubsection{Time-averaged (quenched) input current statistics}\label{sec: input_current_statistics}
      
      Each realization of a spike train is described by a set of spike times $\{ t^{(p)}\}$ that lead to postsynaptic current $I_{i}(t\mid t^{(p)})= J_{kl} f_l(t-t^{(p)})$. The total current to neuron $i$ is
      \begin{equation*}
	    I_{i}(t\mid \{ t^{(p)}\})= J \sum_{l=E,I} \sum_{j=1}^{N_l} A_{kl}^{ij}\sum_{p=1}^{N_{AP}}f_l(t-t^{(p)})\,.
      \end{equation*} 
      Here we denote the total number of spike times $t^{(p)}$ within the interval $t\in [0,T]$ in one compound spike train as $N_{\text{AP}}$. The probability of any realizations of such a spike train with rate $\Omega(t)$ is \cite{dayan_theoretical_2001}
      \begin{equation*}
	    P\left(\{ t^{(p)}\}\mid\Omega(t)\right)=\frac{e^{-\int_{0}^{T}\D t \Omega(t)}}{N_{\text{AP}}!}\prod_{p=1}^{N_{\text{AP}}}\Omega(t^{(p)})\,.
      \end{equation*}
      We obtain the total current into neuron $i$ using the independence of subsequent PSCs, allowing us to integrate over the $N_{\text{AP}}$-dimensional hypercube, covering the temporal course of one spike per dimension. %This  approach allows analytic tractability, however, it neglects the impact of temporally correlated synapses, that could induce synaptic depression.
      
%       It can be decomposed into a mean current $I_0$ averaged over both time and network, quenched fluctuations over the network, described by the variance $\alpha_I^2$ and temporal fluctuations, described by the variance $\sigma_I^2$. The following lines will walk the reader through the calculation of those.
      Accounting for the complete space of possible realization of compound spike trains, we average over all $N_{\text{AP}}\in \mathbb{N}$. Assuming stationarity, the average input current evaluates to (see App.~\ref{sec: appendix_input_current}):
      %\begin{widetext}
      \begin{align}
	    \langle I_k^i(t)\rangle = \frac{1}{\sqrt{K}}\sum_{l=E,I} J_{kl} f_{\mathcal{N}_l} \Omega_l^i\, ,\label{eq: I_tmp_avg}
	    %\Omega \sum_{l=E,I} J_{kl}
      \end{align}
      $f_{\mathcal{N}_l}$ being the integral evaluated over the kernel of population $l$.
      
      %\end{widetext}
    The linear dependence on the compound spike rate allows to determine the mean and quenched variance from the statistics of $\Omega$ from Eq.~\ref{eq: omega_mean} and \ref{eq: omega_var}:
    \begin{align}
	    \bar{I}_k^{\text{rec}} &:= [\langle I_k^{i}(t)\rangle]_i = \sqrt{K} f_{\mathcal{N}_l} (J_{kE}\kappa_E \bar{\nu}_E - J_{kI} \kappa_I \bar{\nu}_I) \label{eq: current_mean}\\
	    \alpha_{I_k}^{2} &:=\left[(\langle I_k^{i}(t)\rangle-\bar{I}_k^{\text{rec}})^{2}\right]_{i}= \sum_{l=E,I} \frac{J_{kl}^2}{K} f_{\mathcal{N}_l}^2\text{Var}\left(\Omega_l^i\right) \nonumber\\
	    &= J_{kE}^2 f_{\mathcal{N}_E}\kappa_E q_E + J_{kI}^2 f_{\mathcal{N}_I} \kappa_I q_I \label{eq: current_var}
    \end{align}
    Note, that the scaling of synaptic weights (Eq.~\ref{eq: syn_weight}) keeps the quenched variance independent of network size. Network- and time-averaged input currents follow a scaling with $K^{\frac{1}{2}}$, while individual neurons' temporal averages of input currents deviate from this according to a Gaussian distribution with variance $\alpha_{I_k}^2$ (c.f. Fig.~\ref{fig: way_to_current}(c)).
    
    \subsubsection{Temporal input current statistics}
    Additional to the neuron-to-neuron differences in quenched input current, we obtained the temporal fluctuations. The statistics of input currents originating from population $l$ can be captured by the covariance function, App.~\ref{sec: appendix_input_current_covariance}:
    \begin{align*}
	    C_{I_{kl}^{i}}(t,t') &:=  \frac{J_{kl}^{2}}{K}\Omega_l^{i}\int_{-\infty}^{\infty}\D s f_l(s-t)f_l(s-t')
    \end{align*}
    Evaluating the integral with the defined PSC-kernel (Eq.~\ref{eq: kernel_single}) and substituting $\Delta t = t-t'$:
    \begin{align}
	    C_{I_{kl}^i}(\Delta t)&= \frac{J_{kl}^2}{K}\Omega_l^i\frac{\tau_{I_l}}{2\mathcal{N}_l^2}\exp\left(-\frac{|\Delta t|}{\tau_{I_l}}\right) \label{eq: input_current_covariance} \\
	    \sigma_{I_{kl}^i}^2=C_{I_{kl}^i}(0)&= \frac{J_{kl}^2 \Omega_l^i \tau_{I_l}}{2 K \mathcal{N}_l^2}
    \end{align}
    The mean and variance of $\sigma_{I_i}$ can be obtained from the compound spike trains statistics (Eq.~\ref{eq: omega_mean},~\ref{eq: omega_var}):
    \begin{align}
	    \sigma_{I_k}^2 &= \sum_{l=E,I} \sigma_{I_{kl}}^2 = \sum_{l=E,I} \kappa_l \frac{J_{kl}^2 \bar{\nu_l} \tau_{I_l}}{2 \mathcal{N}_l}\label{eq: sigma_I}\\ \text{Var}_k^{i}(\sigma_{I_k^i}^2)&=\sum_{l=E,I} \frac{J_{kl}^4 \tau_{I_l}^2}{4 K^2 \mathcal{N}_l^4}\text{Var}(\Omega_l^i) \nonumber \\
	    &= \frac{1}{K} \sum_{l=E,I} \frac{J_{kl}^4 \tau_{I_l}^2}{4\mathcal{N}_l^4}q_l \xrightarrow[K\rightarrow\infty]{} 0 \label{eq: I_var}
    \end{align}
    As expected from the scaling of synaptic weights (Eq.~\ref{eq: syn_weight}), the variance of the temporal fluctuations in large networks is independent of network size ($\text{Var}(\sigma_{I}^2) \rightarrow 0$). All neurons are thus subject to statistically identical temporal fluctuations with variance $\sigma_I^2$. We have thus concluded obtaining all parameters describing the temporal fluctuations, sketched in Fig.~\ref{fig: way_to_current}(d). 
    
    \subsubsection{Overall distribution of input currents}
    
    Summarizing the different internal ($\bar{I}^{rec}$, $\alpha_I$, $\sigma_I$) and external ($I^{ext}$) components of the input current to a single neuron $i$ in population $k$, we obtain:
      %We obtain both variances, $\alpha_I^2$ and $\sigma_I^2$ independent of the network size and - to operate the network in the fluctuation-driven regime that was found to generate a balanced state - it is left to add a constant external current $I^{\text{ext}}$ of the same order ($\mathcal{O}(\sqrt{K})$) of, but with opposite sign to the mean recurrent input current $\bar{I}^{\text{rec}}$. As we chose the neurons to have inhibitory influence, the external current will be excitatory, and thus constantly depolarize all neurons, while it is balanced from the recurrent activity, that hyperpolarizes membrane potentials with each AP. 
    \begin{align}
    	I_k^{i}(t) &= \overbrace{\sqrt{K}(I_k^{\text{ext}} + J_{kE} f_{\mathcal{N}_E} \kappa_E \bar{\nu}_E - J_{kI} f_{\mathcal{N}_I} \kappa_I \bar{\nu}_I)}^{=: \bar{I}_k^0 \text{, network \& time averaged}} \nonumber \\ 
    	&\qquad\qquad\qquad\quad + \underbrace{\alpha_{I_k} x_i}_{\text{network var.}} + \underbrace{\sigma_{I_k} \eta_i(t)}_{\text{temporal var.}}\, , \label{eq: I_total}
    \end{align}
    where $x_i$ is a Gaussian random number with $0$ mean and variance $1$. The temporal course is described by the stochastic function $\eta_i(t)$. % describing a unit variance Ornstein Uhlenbeck (OU) process. %a gaussian random process with the same parameters.
    Dropping the neuron index $i$ for readability, the temporal average $\bar{I}_k = \langle I_k(t)\rangle_t$ is distributed across the network as:
    \begin{equation}
	    \rho(\bar{I}_k) = \frac{1}{\sqrt{2\pi\alpha_{I_k}^{2}}}\exp{\left(-\frac{(\bar{I}_k - \bar{I}_k^0)^2}{2\alpha_{I_k}^{2}}\right)}\label{eq: rho_I}
    \end{equation}

    \subsection*{The balance equations and conditions}
    
    The occurence of a balanced state in these netowrks depends on the parameters $J_{kl}$ and $I_k^{(\text{ext})}$, Eq.~\ref{eq: I_total}.
    
    With balanced inhibitory and excitatory mean inputs, the network dynamics are dominated by the quenched and temporal fluctuations. The high ($\mathcal{O}(\sqrt{K})$) external and recurrent currents $I^{\text{ext}}$ and $\bar{I}^{\text{rec}}$ cancel each other at lowest order in $K$ and determine the balanced state firing rate:
    \begin{equation}
      \mathcal{O}\left(\frac{1}{\sqrt{K}}\right) \approx 0 = I_k^{(ext)} + J_{kE}f_{\mathcal{N}_E}\kappa_E\bar{\nu}_E - J_{kI}f_{\mathcal{N}_I}\kappa_I\bar{\nu}_I\;.\label{eq: balanceI}
    \end{equation}
    Self-consistent solutions with two active populations can only be achieved when Eq.~\ref{eq: balanceI} holds for both populations and results in: 
    \begin{align}
      \bar{\nu}_E^{\text{bal}}=\bar{\nu}_E&=\frac{1}{f_{\mathcal{N}_E}\kappa_E}\frac{I_E^{(ext)}J_{II} - I_I^{(ext)}J_{EI}}{J_{EI}J_{IE} - J_{EE}J_{II}} + \mathcal{O}\left(\frac{1}{\sqrt{K}}\right)\label{eq: nu_balance_E}\\
      \bar{\nu}_I^{\text{bal}}=\bar{\nu}_I&=\frac{1}{f_{\mathcal{N}_I}\kappa_I}\frac{I_E^{(ext)}J_{IE} - I_I^{(ext)}J_{EE}}{J_{EI}J_{IE} - J_{EE}J_{II}} + \mathcal{O}\left(\frac{1}{\sqrt{K}}\right) \label{eq: nu_balance_I}
    \end{align}
    In the large network limit ($K\rightarrow \infty$), these balance equations ensure the average input currents $\bar{I}_k^0$ to remain bounded, while the $\mathcal{O}\left(K^{-\frac{1}{2}}\right)$-deviation of the balanced firing rate $\bar{\nu}^{\text{bal}}$ is responsible for a small, non-vanishing deviation in finite size networks.
    
    Contrary to the single-population case ($\bar{\nu}^{\text{bal}} = \frac{I^{\text{ext}}}{J} + \mathcal{O}\left(\frac{1}{\sqrt{K}}\right)$), the Balanced State in two-population networks can not be obtained for all synaptic weights, as some will lead to exploding ($\bar{\nu}_k \rightarrow \infty$) or quiescent ($\bar{\nu}_k = 0\,$Hz) solutions to the firing rate distributions.

  \subsubsection*{External driving currents}
    
    The population of excitatory neurons is the only internal source of activity in the network. Even though a state of self-sustained dynamics without an external driving current ($I_E^{\text{ext}}=0$) is possible when inhibitory influence is weak, we will exclude this to focus on models that describe cortical areas driven by afferent, long-range excitatory currents to process and forward information. We therefore require 
    \begin{equation}
      I_E^{\text{ext}} > 0\,.\label{eq: I_E_pos}
    \end{equation}
    For the inhibitory population, on the other hand, we will allow for a vanishing external drive, thus $I_I^{\text{ext}} \geq 0$. This requires the excitatory population to project onto the inhibitory population with non-vanishing synaptic weights $J_{IE} > 0$, to avoid quiescent solutions.

%     This allows for two different network setups: one with an external driving current projecting to both populations and one that only reaches towards the excitatory one, while the inhibitory population is driven by the spiking activity of its excitatory counterpart, only.
    
  \subsubsection*{Synaptic weights}
    Eq.~\ref{eq: balanceI} states that the mean input from the inhibitory network not only has to cancel the input from the excitatory neurons, but also the external drive. Thus we obtain:
    \begin{align}
      J_{EE}f_{\mathcal{N}_E}\kappa_E\bar{\nu_E} &< J_{EI}f_{\mathcal{N}_I}\kappa_I\bar{\nu_I}\; ,\label{eq: cons_ext_1}\\ J_{IE}f_{\mathcal{N}_E}\kappa_E\bar{\nu_E} &\leq J_{II}f_{\mathcal{N}_I}\kappa_I\bar{\nu_I}\; ,\label{eq: cons_ext_2}
    \end{align}
    where the latter implements the case of a non-driven inhibitory population. Further demanding the inhibitory synapses to have larger influence than the excitatory ones, excludes solutions of exploding excitatory activity from the network dynamics, as the inhibitory feedback can balance the overall excitation for all values:
    \begin{equation}
      J_{EE} f_{\mathcal{N}_E}\kappa_E < J_{EI} f_{\mathcal{N}_I}\kappa_I \;, \quad J_{IE} f_{\mathcal{N}_E}\kappa_E < J_{II} f_{\mathcal{N}_I}\kappa_I \label{eq: non_explo}
    \end{equation}
    The case of exploding inhibitory activity is excluded by the balanced state dynamics, in which excess inhibition leads to a shutdown of activity and thus prohibits itself.
    
    While quiescent solutions for the inhibitory population are impossible, as long as it receives any input, we exclude quiescent excitatory populations explicitly by demanding 
    \begin{align}
      &\frac{J_{EI}J_{IE}}{J_{EE}J_{II}}>1 , \\ 
      &I_E^{(ext)}J_{II} > I_I^{(ext)}J_{EI} \, , \qquad I_E^{(ext)}J_{IE} > I_I^{(ext)}J_{EE}\,,
    \end{align}
    which can be found by solving eq.~\ref{eq: nu_balance_E},~\ref{eq: nu_balance_I} for $\bar{\nu}_E = 0$. Summarizing this and eq.~\ref{eq: cons_ext_1},~\ref{eq: cons_ext_2},~\ref{eq: non_explo}, we obtain Balanced State networks with constraints:
    
    \begin{align}
      &\frac{J_{EE}}{J_{EI}} < \frac{J_{IE}}{J_{II}} \leq \frac{f_{\mathcal{N}_I}\kappa_I}{f_{\mathcal{N}_E}\kappa_E}\cdot\text{min}\left\{1,\frac{\bar{\nu_I}}{\bar{\nu_E}}\right\}\qquad \text{and} \nonumber \\
      &\frac{J_{EE}}{J_{IE}} < \frac{J_{EI}}{J_{II}} < \frac{I_E^{(ext)}}{I_I^{(ext)}} \label{eq: weightbounds}
    \end{align}
    To achieve comparable statistics of network dynamics between two-population and one-population models, we can furthermore require the temporal variance of input currents to be equal in both populations ($\sigma_E^2$, $\sigma_I^2$) and thus conveniently compared to a one-population model ($\sigma_{I^{(1)}}^2$, inhibitory population only), in the case of same firing rates ($\bar{\nu}_E=\bar{\nu}_I=\bar{\nu}_{I^{(1)}}$) and same synaptic time constants ($\tau_{I_E} = \tau_{I_I} = \tau_{I^{(1)}}$):
    \begin{align}
      &\sigma_E^2 = \sigma_I^2 = \sigma_{I^{(1)}}^2 \\
      &\Leftrightarrow \frac{J_{EE}^2 \bar{\nu}_E}{2\tau_{I_E}} + \frac{J_{EI}^2 \bar{\nu}_I}{2\tau_{I_I}} = \frac{J_{IE}^2 \bar{\nu}_E}{2\tau_{I_E}} + \frac{J_{II}^2 \bar{\nu}_I}{2\tau_{I_I}} = \frac{J^2 \bar{\nu}_{I^{(1)}}}{2\tau_{I^{(1)}}}\\
      &\Leftrightarrow J_{EE}^2 + J_{EI}^2 = J_{IE}^2 + J_{II}^2 = J^2
    \end{align}
    We introduce two parameters that reduce the dimensionality of the problem and fulfill the discussed constraints:
    \begin{align}
      \begin{pmatrix} J_{EE} & J_{EI} \\ J_{IE} & J_{II} \end{pmatrix} &= \begin{pmatrix} \eta \varepsilon & \sqrt{1-(\eta \varepsilon)^2} \nonumber\\ \varepsilon & \sqrt{1-\varepsilon^2} \end{pmatrix}\; , \\
      &\eta < 1\;,\quad \varepsilon < \sqrt{\frac{1}{2}} \label{eq: weight}
    \end{align}
    The excitatory-inhibitory feedback loop strength, $\varepsilon$, defines the impact both populations have on the whole network and the ratio of inter-population excitatory coupling $\eta = \frac{J_{EE}}{J_{IE}}$ describes the difference between the influence of both populations on the excitatory neurons.

%         \textbf{get citations from hiemeyer!}
  \subsection{Neuron model}
    
    We utilize the analytically highly tractable Gauss-Rice neuron model \cite{jung_stochastic_1995,naundorf_unique_2006,rice_mathematical_1944} which follows the dynamics of a leaky integrate-and-fire (LIF) neuron
    \begin{equation}
	    \tau_M\dot{V}(t)=-(V(t)-V_R)+I(t) \, , \label{eq: gov}
    \end{equation}
    that generates an AP when the membrane potential $V(t)$ reaches a threshold $\Psi_0$ from below, but does not include a reset to some baseline potential at spike time. The temporal scale is set by the membrane time constant $\tau_M$. For the sake of readability, the population index $k$ is dropped until it becomes relevant to the calculations.
      
%       The calculations will use the neural dynamics to compute the membrane voltage statistics arising from a presynaptic current with mean value $I$. As the temporal fluctuations of $I$ are the same in every neuron of our network (Eq.~\ref{eq: I_var}), the calculations will hold for all neurons and we will therefore omit the according indices. Knowing the statistics of the membrane potential, an analytic approach to spike counting will allow us to infer the resulting firing rate and thus the shape of the f-I-curve.
    
    Without loss of generality, we set $V_R = 0$. Following Jung et al \cite{jung_stochastic_1995} the rate of AP initiation is obtained from counting the number of events at which the potential crosses the threshold value ($V(t)=\Psi_0$) from below ($\dot{V}(t)>0$): 
    \begin{equation}
	    \nu =\langle\delta\left(V(t)-\Psi_0\right)\Theta(\dot{V}(t))\dot{V}(t)\rangle_t \,. \label{eq: spike_ct}
    \end{equation}
    where $\delta(\cdot)$ and $\Theta(\cdot)$ denote the Dirac-Delta-function and Heavyside-function, respectively. Due to ergodicity, the temporal average $\langle \cdot \rangle$ can be expressed by the joint probability density function of the membrane potential and its derivative
    \begin{equation}
	    \nu=\int_{-\infty}^{\infty}{\int_{-\infty}^{\infty}{\D V \,\D \dot{V}\,P(V,\dot{V})\,\delta(V-\Psi_0)\,\Theta(\dot{V})\,\dot{V} }} \label{eq: f-I_raw}
    \end{equation}
    Hence, the statistics of $V$ and $\dot{V}$ are sufficient to obtain the firing rate of the neuron model used here.
    
    \subsubsection{Membrane potential statistics}
        The joint distribution $P(V,\dot{V})$ - a multidimensional Gaussian, as its variables are driven by a Gaussian distributed input current - is fully characterized by the variables' mean values $\left\langle V \right\rangle$, $\left\langle \dot{V} \right\rangle$ and the correlation matrix
        %As the membrane potential dynamics are governed by the Gaussian distributed incoming currents, their dynamics obey Gaussian statistics as well, thus further require only the variance for a complete description.
        \begin{align}
            C = \left(\begin{array}{cc}
                 \left\langle V^2 \right\rangle & \left\langle V \dot{V}  \right\rangle \\
                 \left\langle V \dot{V} \right\rangle & \left\langle \dot{V}^2 \right\rangle
            \end{array}\right) \label{eq: methods_V_correlation_matrix}
        \end{align}
        Mean values $\langle V(t) \rangle_t = \langle I(t) \rangle$ and $\langle \dot{V}(t) \rangle_t = 0$ can be extracted directly from Eq.~\ref{eq: gov} and the constraint of a stationary solution. The assumption of stationarity further leads to a vanishing cross-correlation, implying independence as they are jointly normal distributed:
        \begin{align*}
	        \left\langle V(t)\dot{V}(t) \right\rangle &= \left\langle \frac{1}{2}\frac{\D}{\D t}V^2(t)\right\rangle=\frac{1}{2}\frac{\D}{\D t}\left\langle V^2(t)\right\rangle \\
	        &= \frac{1}{2}\frac{\D}{\D t}(\sigma_V^2 + \bar{V}^2)=0\;.
        \end{align*}
        where we used the definition of the variance $\sigma_V^2=\left\langle V^2(t)\right\rangle-\left\langle V(t)\right\rangle^2$. $P(V,\dot{V})$ (Eq.~\ref{eq: f-I_raw}) thus factorizes into the Gaussian pdfs $P(V)$ and $P(\dot{V})$, for which only the respective variances $\sigma_V^2$, $\sigma^2_{\dot{V}}$ are left to obtain for a complete characterization:
        
        Within a neuron, an incoming current pulse $\delta I(t)$ will lead to a temporally extended change in membrane potential according to
        \begin{equation*}
            \delta V(t)=\int_{-\infty}^{+\infty}\D s G(t-s)\delta I(s)\,,
        \end{equation*}
        with $G(\Delta t) = \frac{1}{\tau_M}\exp\left(-\frac{\Delta t}{\tau_M}\right)\Theta(\Delta t)$ the Greens function, obtained from Eq.~\ref{eq: gov}. 
        
        The contributions from both populations input currents are additive, $C_{V_k}(\Delta t) = \sum_{l=E,I} C_{V_{kl}}$, allowing to evaluate the integral separately for each population, thus having distinct contributions from each population $l$. The distinction between input populations $l$ further motivates the reintroduction of the population indices.
        
        Evaluation of the first entry of the correlation matrix Eq.~\ref{eq: methods_V_correlation_matrix} for the contribution of population $l$ results in (see App.~\ref{sec: appendix_potential_covariance} for detailed calculations and the solution in the case $\tau_{I_l} = \tau_M$):
        \begin{align}
            &C_{V_{kl}}(\Delta t) = \frac{\sigma_{I_l}^{2}\tau_{I_l}}{\tau_{I_l}^{2}-\tau_M^{2}}\left[\tau_{I_l} e^{-\frac{\left|\Delta t \right|}{\tau_{I_l}}}-\tau_M e^{-\frac{\left|\Delta t \right|}{\tau_M}}\right] %&,\,\text{for } \tau_{I_l}\neq \tau_M\\
                %&\\
                %\frac{\sigma_{I_l}^{2}}{2}\left(1+\frac{\left|\Delta t\right|}{\tau_M}\right)\exp\left(-\frac{\left|\Delta t \right|}{\tau_M}\right)&,\,\text{for } \tau_{I_l}=\tau_M\end{array} \label{eq: V_corr}
            %\right.
        \end{align}
        The variance is obtained from $\sigma_{V_k}^2 = C_{V_k}(0)$, while the variance of the membrane potential derivative is obtained as $\sigma^2_{\dot{V_k}} = \left.\frac{\D^2}{\D t^2} C_{V_k}(\Delta t)\right|_{\Delta t = 0}$:
        \begin{align}
            \sigma_{V_k}^2 &= \sum_{l=E,I}{\sigma_{V_{kl}}^2}\,,\qquad \sigma_{V_{kl}}^2=C_{V_{kl}}(0)=\frac{\sigma_{I_{kl}}^2\tau_{I_l}}{\tau_{I_l}+\tau_M}\\
            \sigma_{\dot{V}_k}^2 &= \sum_{l=E,I}{\sigma_{\dot{V}_{kl}}^2}\,,\qquad \sigma_{\dot{V}_{kl}}^2=C_{\dot{V}_{kl}}(0)=\frac{\sigma_{V_{kl}}^2}{\tau_{I_l}\tau_M}\;.\label{eq: sigma_V}
        \end{align}
%       As expected, the correlation functions depend on the synaptic time constant $\tau_I$.
        
        The limiting case $\tau_{I_l}=0$ would correspond to synapses coupled by $\delta$-pulses that result in a white noise current and a membrane potential following an OU process. In this limit, however, $\sigma_{\dot{V_k}}$ is not finite. % (Fig.~\ref{fig: corr_tau}). %In neurons with a reset after AP initiation this limiting case offers no problems, however the Gauss-Rice approach requires a finite current decay rate to be applicable.
        
        This concludes the derivation of membrane potential pdfs:
        \begin{align*}
            P(V_k,\dot{V_k}) &= P(V_k) P(\dot{V_k}) \\
            P(V_k) &= \frac{1}{\sqrt{2\pi}\sigma_{V_k}} \exp{\left( -\frac{(V_k-\bar{V_k})^2}{2\sigma_{V_k}^2}\right)}\\
            P(\dot{V_k}) &= \frac{1}{\sqrt{2\pi}\sigma_{\dot{V}_k}} \exp{\left( -\frac{\dot{V_k}^2}{2\sigma_{\dot{V}_k}^2}\right)}\,
        \end{align*}
        and allows for the evaluation of the input-current depending firing rate (Eq.~\ref{eq: f-I_raw}), using the unitlessness of Eq.~\ref{eq: gov} to find $\bar{V}_k = \bar{I}_k$, see also \cite{tchumatchenko_signatures_2010}:
        \begin{align}
    	    \nu_k(\bar{I}_k^i,\sigma_{V_k},\sigma_{\dot{V}_k}) &= \nu_k^{\text{max}}(\sigma_{V_k},\sigma_{\dot{V}_k})\exp\left(-\frac{(\bar{I}_k^i-\Psi_0)^{2}}{2\sigma_{V_k}^{2}}\right)\; ,\\
    	    \nu_k^{\text{max}}(\sigma_{V_k},\sigma_{\dot{V}_k}) &= \frac{1}{2\pi} \frac{\sigma_{\dot{V}_k}}{\sigma_{V_k}}\, .\label{eq: f-I}
        \end{align}
        This is the firing rate transfer function, which translates any Gaussian statistics of input currents to the activity response of a neuron. It takes on a Gaussian shape with maximum firing rate $\nu^{\text{max}}$ which has a linear dependence on the ratio of variances $\frac{\sigma_{\dot{V}}}{\sigma_V}$. This  simplifies to $(\tau_{I_l}\tau_M)^{-\frac{1}{2}}$ in the case of a single, inhibitory population. In the case of average input currents $\bar{I}_k^i\geq\Psi_0$ the Gauss-Rice model loses plausibility due to the missing reset after threshold crossing. This condition, however, is never met as long as low firing rates are maintained.

    \subsection{Mixed receptor types}\label{sec: methods_mixed_kernel}
        So far we considered the case of a single synaptic time-constants, but the model remains highly tractable also for a mixture of post-synaptic receptors, Eq.~\ref{eq: kernel}. We thus turn to the case of a mixed excitatory post-synaptic kernel, comprised of fast AMPA and slow NMDA receptors. The mixing parameters $r_m$ are associated through $r_{AMPA} = 1-r_{NMDA}$, henceforth $r_{NMDA}=r$.
  
        The resulting correlation function of the excitatory input current then evaluates to
        \begin{align*}
            C_{I_k}(t)&=J_{kE}^2\kappa_E\bar{\nu}_E \int_{-\infty}^{\infty}{f_E(t-s)f_E(-s)\D s}\\
%     &= (1-r)^2 C_A(t) + r^2 C_N(t) + (1-r) r\frac{J_{k,E}^2\kappa_E\bar{\nu}_E+J_{k,I}^2\kappa_I\bar{\nu}_I}{\tau_A\tau_N}\int_{-\infty}^{\infty}{\left[\exp{\left(\frac{s}{\tau_A}+\frac{s-t}{\tau_N}\right)}+\exp{\left(\frac{s}{\tau_N}+\frac{s-t}{\tau_A}\right)}\right]\Theta(-s)\Theta(t-s)\D s}\\
            &= (1-r)^2 C_A(t) + r^2 C_N(t) \\
            + &(1-r)\, r\frac{J_{kE}^2\kappa_E\bar{\nu}_E}{\tau_A+\tau_N}\left[\exp{\left(-\frac{|t|}{\tau_A}\right)}+\exp{\left(-\frac{|t|}{\tau_N}\right)}\right]\;,
        \end{align*}
        where $C_A(t)$ denotes the excitatory correlation function obtained from a network of neurons with only AMPA-receptors and $C_N(t)$ is the excitatory correlation function of a network of neurons with only NMDA-receptors, Eq.~\ref{eq: input_current_covariance}. The correlation function of the membrane potential from the excitatory input again can be calculated according to Appendix~\ref{sec: appendix_potential_covariance}:
        \begin{widetext}
        \begin{align*}
            C_{V_{kE}}(t)=\int_{-\infty}^{\infty}{ C_{I_k}(s)G^{(2)}(s-t)\D s}%\\
            %&=(1-r)^2 C_{V,A}(t) + r^2 C_{V,N}(t) + (1-r) r\frac{J_{k,E}^2\kappa_E\bar{\nu}_E+J_{k,I}^2\kappa_I\bar{\nu}_I}{(\tau_A+\tau_N)\tau_M}\int_{-\infty}^{\infty}{\left[\exp{\left(-\frac{|t|}{\tau_A}\right)}+\exp{\left(-\frac{|t|}{\tau_N}\right)}\right]\exp{\left(-\frac{|s-t|}{\tau_M}\right)}\D s}\\
            &=\frac{J_{kE}^2\kappa_E\bar{\nu}_E}{\tau_A^2-\tau_M^2}\left[\frac{(1-r)^2}{2}+\frac{(1-r)\,r\,\tau_A}{\tau_A + \tau_N}\right]\left[\tau_A\exp{\left(-\frac{|t|}{\tau_A}\right)}-\tau_M\exp{\left(-\frac{|t|}{\tau_M}\right)}\right]\\
            &\quad +\frac{J_{kE}^2\kappa_E\bar{\nu}_E}{\tau_N^2-\tau_M^2}\left[\frac{r^2}{2}+\frac{(1-r)\,r\,\tau_N}{\tau_A + \tau_N}\right]\left[\tau_N\exp{\left(-\frac{|t|}{\tau_N}\right)}-\tau_M\exp{\left(-\frac{|t|}{\tau_M}\right)}\right]\;.
        \end{align*}
        \end{widetext}
        The variance follows as
        \begin{align}
            \sigma_{V_{kE}}^2=C_{V_k}(0)=\overbrace{\frac{J_{kE}^2\kappa_E\bar{\nu}_E}{\tau_A +\tau_M}\left[\frac{(1-r)^2}{2}+\frac{(1-r)\,r\,\tau_A}{\tau_A + \tau_N}\right]}^{=: \sigma_{V,A}^2} \nonumber \\
            +\underbrace{\frac{J_{kE}^2\kappa_E\bar{\nu}_E}{\tau_N +\tau_M}\left[\frac{r^2}{2}+\frac{(1-r)\,r\,\tau_N}{\tau_A + \tau_N}\right]}_{=: \sigma_{V,N}^2}\;.\label{eq: mixedvariance}
        \end{align}
        We obtain the variance of the derivative of the membrane potential, originating from excitatory sources as:
        \begin{align}
            \sigma_{\dot{V}_{kE}}^2 &= \frac{\partial^2 C_{V_k}(0)}{\partial t^2}\\
            %&=\frac{J_{kE}^2\kappa_E\bar{\nu}_E}{\tau_A^2 -\tau_M^2}\left[\frac{(1-r)^2}{2}+\frac{(1-r)\,r\,\tau_A}{\tau_A + \tau_N}\right]\left[\frac{1}{\tau_A} -\frac{1}{\tau_M}\right] + \frac{J_{kE}^2\kappa_E\bar{\nu}_E}{\tau_N^2 -\tau_M^2}\left[\frac{r^2}{2}+\frac{(1-r)\,r\,\tau_N}{\tau_A + \tau_N}\right]\left[\frac{1}{\tau_N} -\frac{1}{\tau_M}\right]\nonumber\\
            \hspace*{0.2cm}
            &= -\frac{1}{\tau_A\tau_M}\frac{J_{kE}^2\kappa_E\bar{\nu}_E}{(\tau_A +\tau_M)}\left[\frac{(1-r)^2}{2}+\frac{(1-r)\,r\,\tau_A}{\tau_A + \tau_N}\right] \nonumber \\
            &\quad-\frac{1}{\tau_N\tau_M}\frac{J_{kE}^2\kappa_E\bar{\nu}_E}{(\tau_N +\tau_M)}\left[\frac{r^2}{2}+\frac{(1-r)\,r\,\tau_N}{\tau_A + \tau_N}\right]\nonumber\\
            &= -\frac{\sigma_{V,A}^2}{\tau_A\tau_M} - \frac{\sigma_{V,N}^2}{\tau_N\tau_M}\;.\label{eq: dervarn}
        \end{align}
        %These results allow us to calculate the networks statistics just as shown before, by substituting the temporal variances throughout all equations with the above results for mixed receptor types.
    
    \subsection{Introducing network heterogeneity}\label{sec: methods_heterogeneity}
        Following \cite{van_vreeswijk_chaos_1996,van_vreeswijk_chaotic_1998}, we include a heterogeneity $\alpha_{0,k}$ in the firing threshold of population $k$:
        \begin{equation*}
	        \Psi_{0,k}\rightarrow \Psi_0 + \alpha_{0,k} \cdot x\,
        \end{equation*}
        where $x$ is a Gaussian random number with mean $0$ and unit variance. Note that, in a linear neuron model such as ours, a higher firing threshold of a neuron is equivalent to a lower received average input current, as both result in an additional depolarization required for emitting an action potential. Therefore the total quenched variance including Gaussian threshold differences between individual neurons and contributions from the random connectivity is defined as:
        \begin{equation*}
	        \alpha_k^2 := \alpha_{I_k}^2 + \alpha_{0,k}^2\; , \qquad q_k \; \rightarrow \; q_k+q_{0,k} \, ,
        \end{equation*}
        where $q_{0,k} = \alpha_{0,k}^2/J_{kl}^2$.
        
        %As the threshold inhomogeneity is independent of the network firing rate and the quenched variance scales with the second moment $q$, we expect a huge influence of non-zero values of $\alpha_0$ at low firing rates, which will be confirmed in the further analysis.
        
    \subsection{Firing rate distributions}\label{sec: methods_firingrate}
    \begin{figure}
      \centering
      \includegraphics[width=\linewidth]{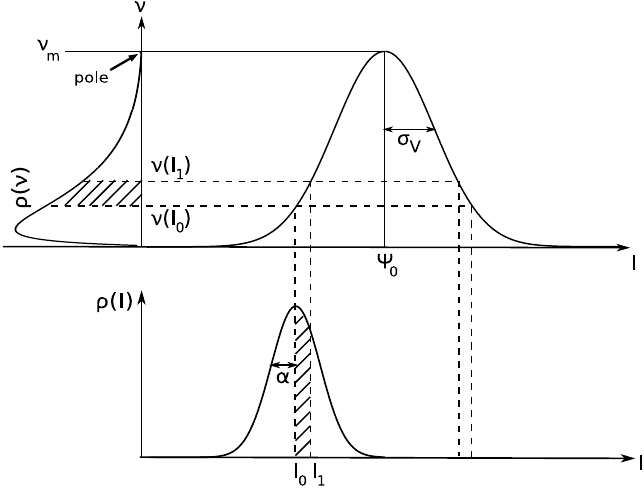}
      \caption{The firing rate distribution is obtained by a transformation from the distribution of mean input currents $\rho(I)$ (lower graph) through the transfer function of the Gauss-Rice neuron $\nu(I)$ (upper right graph); left: resulting firing rate $\rho(\nu)$. Dashed lines sketch, how probability mass is conserved under the transformation. Once the two former distributions are known the firing rate distribution can be determined. $\nu_m = \bar{\nu}^{\text{max}}$.}
      \label{fig: way_to_distr}
    \end{figure}
    
%     The preceding analytics evaluated the input current statistics in a sparse network in the balanced state by finding the mean and variance. According to the central limit theorem, those statistics are sufficient and yield the probability distribution of recurrent currents in the network. Even though the total input current $\bar{I}_0$ is still unknown, t
    The distribution of average input currents $\rho(I)$ (Eq.~\ref{eq: rho_I}) together with the firing rate response of a neuron to a mean input current $\nu(I)$ (Eq.~\ref{eq: f-I}) allows for the calculation of the distribution of average firing rates across each population $k$ (population index dropped), using a transformation of probability density functions (see Fig.~\ref{fig: way_to_distr}) as:
    \begin{equation}
      \rho(\nu) = \rho\left(I_{-}(\nu)\right)\left|\frac{dI_{-}}{d\nu}\right| + \rho\left(I_{+}(\nu)\right)\left|\frac{dI_{+}}{d\nu}\right|\, , \label{eq: transform}
    \end{equation}
    where $-$ and $+$ are the indices to the rising and falling (non-realistic) part of the transfer function, respectively. %To be exact, we include both parts in the calculations, even though the falling part of the f-I-curve is not realistic, as the distribution of input currents has a negligible contribution for fluctuation driven neurons.
    The inverse of Eq.~\ref{eq: f-I} gives $I$ and its absolute derivative for both parts of the transfer function:
    \begin{align*}
        \bar{I}_{\pm}(\nu)&=\Psi_{0}\pm\sigma_{V}\sqrt{-2\ln\frac{\nu}{\nu^{\text{max}}}}\\
        \left|\frac{\D I_{\pm}}{\D \nu}\right|&=\frac{\sigma_{V}}{\nu\sqrt{-2\ln\frac{\nu}{\nu^{\text{max}}}}}
    \end{align*}
    Evaluating Eq.~\ref{eq: transform} and using $\exp(x)+\exp(-x)\equiv2\cosh(x)$, the analytical form of the firing rate distribution is:
%       The two branches of Eq.~\ref{eq: transform} thus give \begin{align*}
%       \rho\left(I_{\pm}(\nu)\right) = & \frac{1}{\sqrt{2\pi}\alpha_I}\exp\left(-\frac{\left(\Psi_{0}\pm\sigma_{V}\sqrt{-2\ln\frac{\nu}{\nu_{\text{max}}}}-\bar{I}\right)^{2}}{2\alpha_I^{2}}\right)\\
% 	= & \frac{1}{\sqrt{2\pi}\alpha_I}\exp\left(-\frac{(\Psi_0-\bar{I})^2}{2\alpha_I^2}\right)\left(\frac{\nu}{\nu_{\text{max}}}\right)^{\frac{\sigma_V^2}{\alpha_I^2}}\exp\left(\pm\frac{(\Psi_0-\bar{I})\sigma_V\sqrt{-2\ln\frac{\nu}{\nu_{\text{max}}}}}{2\alpha_I^2}\right)\,.
% 	\end{align*}
%       Evaluating Eq.~\ref{eq: transform} by summing up the two branches with differing signs in the exponentials and using , we arrive at 
    \begin{align}
        \rho(\nu)=\frac{\gamma}{\nu^{\text{max}}\sqrt{-\pi\ln\frac{\nu}{\nu^{\text{max}}}}}\exp\left(-\frac{\delta^2}{2}\right)\left(\frac{\nu}{\nu^{\text{max}}}\right)^{\left(\gamma^2-1\right)} \cdot \nonumber\\ \cosh\left(\gamma\delta\sqrt{-2\ln\frac{\nu}{\nu^{\text{max}}}}\right) \label{eq: distribution}\\
        \text{with }\quad \gamma^2:=\frac{\sigma_{V}^2}{\alpha^2}\quad,\qquad\delta:=\frac{\Psi_{0}-\bar{I}_0}{\alpha}\,.\nonumber
    \end{align}
    %and $\alpha^2$ the quenched variance across the network. In a homogeneous network it is the same as the variance of input currents $\alpha_I^2$. 
    
    The log firing rate-distribution can be used for further analysis of extrema and evaluates to
    \begin{align}
        \ln \rho(\nu) &=
        -\ln\left(\frac{\nu^{\text{max}}}{\gamma}\sqrt{-\pi\ln\frac{\nu}{\nu^{\text{max}}}}\right)-\frac{\delta^2}{2}+\nonumber\\
        &\quad\left(\gamma^2-1\right)\ln\frac{\nu}{\nu^{\text{max}}}+\ln\left( \cosh{\gamma\delta\sqrt{-2\ln\frac{\nu}{\nu^{\text{max}}}}}\right)\nonumber\\
        &\approx \sqrt{-\ln\frac{\nu}{\nu^{\text{max}}}}\left(\gamma\delta\sqrt{2}-\left(\gamma^{2}-1\right)\sqrt{-\ln\frac{\nu}{\nu^{\text{max}}}}\right)\,\label{eq: distribution_log}
    \end{align}
    where the approximation on the last line holds in the low firing rate limit $\nu\rightarrow 0^+$. It follows from rewriting $2\cosh(\xi)=\exp(\xi)+\exp(-\xi)\approx\exp(\xi)$ and identifies the dominant term $\xi=\gamma\delta\sqrt{-2\ln\frac{\nu}{\nu^{\text{max}}}}$, as $\lim_{x\rightarrow 0}\ln(x) \rightarrow -\infty$.

    \subsubsection*{Peaked distributions}
        
        The exact logarithmic distribution can further be used to obtain the firing rate peak position by finding $\nu$ at which $\frac{\D}{\D \nu} \ln \rho(\nu) = 0$, $\frac{\D^2}{\D\nu^2} \ln \rho(\nu) < 0$, resulting in
        \begin{align}
	        \nu^{(p)}&=\nu^{\text{max}} e^{-\frac{\gamma^{2}\delta^{2}-2\left(\gamma^{2}-1\right)+\gamma\delta\sqrt{\gamma^{2}\delta^{2}-4\left(\gamma^{2}-1\right)}}{4\left(\gamma^{2}-1\right)^{2}}}\label{eq: nu_peak}\\
 	        &\Rightarrow 4\left(\gamma^{2}-1\right)<\gamma^{2}\delta^{2}\,,\label{eq: bound_no_peak}
	    \end{align}
	    with Eq.~\ref{eq: bound_no_peak} defining the requirement for a peaked distribution to occur at all. If it is violated, the firing rate distribution monotonously rises from $\rho(\nu=0) = 0$ towards $\rho(\nu = \nu^{\text{max}}) \rightarrow \infty$. Equality of the terms defines the firing rate $\nu_{\text{no peak}}$, at which a transition between peaked and non-peaked distributions occurs.
	    
	    To measure the skewness of the distribution, we define the \textit{skewness coefficient} $\chi$ as the log-ratio of peak position $\nu^{(p)}$ to mean firing rate $\bar{\nu}$:
    	\begin{equation}
    	  \chi=-\log_{10}\frac{\nu^{(p)}}{\bar{\nu}} \label{eq: definition_chi}
    	\end{equation}
    	Different values of $\chi$ describe, how strongly the distribution leans to one side. A positive value corresponds to right-skewed firing rate distributions, where the majority of the probability mass accumulates at firing rates lower than the average. For $\chi=0$ the peak coincides with the mean rate, for $\chi=1$ the peak occurs at one tenth of the mean, etc. 

%    \clearpage
    \section{Results}
    In this work, we present a model of a neuronal network for which we derive an analytically solvable solution to the firing rate distribution (Methods ~\ref{sec: methods}). The model is able to implement firing threshold heterogeneities, can describe two interconnected neuronal populations, with the ability to add more and includes the possibility for population-specific mixed post-synaptic receptor types.
    
    %To understand how pathological distributions of firing rates can emerge in a neural network, including neurons that emit action potentials (APs) at extremely high rates, as well as a large population of virtually silent neurons, 
    
    We use the analytically tractable Gauss-Rice neuron model in a balanced state network to describe distributions of recurrent input currents and according membrane potentials, from which we derive the firing rate distribution for population $k$:
    \begin{widetext}
    \begin{equation}
        \rho(\nu_k^i)=\frac{\gamma_k}{\nu_k^{\text{max}}\sqrt{-\pi \ln{\frac{\nu_k^i}{\nu_k^{\text{max}}}}}}\exp{\left(-\frac{\delta_k^2}{2}\right)}\left(\frac{\nu_k^i}{\nu_k^{\text{max}}}\right)^{\gamma_k^2 - 1} \cosh{\left(\gamma_k \delta_k \sqrt{-2\ln{\frac{\nu_k^i}{\nu_k^{\text{max}}}}}\right)}\;\text{, with }\quad \delta_k:=\frac{\Psi_0-\bar{I}_k}{\alpha_k} \; , \quad \gamma_k := \frac{\sigma_{V_k}}{\alpha_k}\label{eq: rate_distribution_Gauss}
    %    \rho(\nu)=\frac{\gamma}{\nu_{\text{max}}\sqrt{-\pi\ln\frac{\nu}{\nu_{\text{max}}}}}\exp\left(-\frac{\delta^2}{2}\right)\left(\frac{\nu}{\nu_{\text{max}}}\right)^{\left(\gamma^2-1\right)}\cosh\left(\gamma\delta\sqrt{-2\ln\frac{\nu}{\nu_{\text{max}}}}\right)\,, \quad \text{with}\quad \gamma^2:=\frac{\sigma_{V}^2}{\alpha^2} \, ,\quad \delta:=\frac{\Psi_{0}-\bar{I}_0}{\alpha} \label{eq: rate_distribution_Gauss}
    \end{equation}
    \end{widetext}
    with the \textit{dark matter exponent} $\gamma$, the ratio of temporal variance of membrane potential fluctuation $\sigma_V^2$ and the quenched variance of input currents across the network $\alpha^2$ and the relative distance $\delta$ of mean input current $I_0$ to the firing threshold $\Psi_0$.
    
    %The model is entirely characterised by the synaptic weights and four additional parameters per population $l$: the average firing rate $\bar{\nu}_l$ as a measure for the general activity level; the synaptic and membrane time constants $\tau_{I_l}$ and $tau_M$, respectively, describing the timescales over which action potentials are transmitted and processed; the synaptic weight $J_0$, here assumed to be a network constant; and finally the added heterogeneity $\alpha_0$, allowing for the solution of homogeneous, as well as heterogeneous networks, accumulating assumed Gaussian distributed heterogeneities in $\alpha^2$ additional to the heterogeneities imposed by randomly distributed in-degrees of the Erd\"{o}s-R\'{e}nyi connectivity. % $\delta:=\frac{\Psi_{0}-\bar{I}_0}{\alpha}$ describes the distance of the mean input current $I_0$ to the firing threshold $\Psi_0$ in multiples of $\alpha$.
  
    %The quenched variance $\alpha^2$ of input currents not only contains network effects, but also includes potential heterogeneities, e.g. in the firing threshold, distributed as $\mathcal{N}(0,\alpha_0)$ (Methods~\ref{sec: methods_heterogeneity}).
  
    \subsection{Selfconsistent solutions to the firing rate distribution}
        Eq.~\ref{eq: rate_distribution_Gauss} depends on both, the first moment (directly via $\bar{\nu}_l:=\left[\nu_l^i\right]_i$, indirectly via the temporal variance $\sigma_{V_l}^2$, Eq.~\ref{eq: sigma_V}) and the second moment (with $\alpha_{I_{kl}}^2 =  J_{kl}^2 f_{\mathcal{N}_l}\kappa_l q_l$; $q_l:=\left[ (\nu_l^i)^2 \right]_i$, Eq.~\ref{eq: current_var}) of the firing rates $\nu_l$. These moments can be calculated population-wise from the distribution of input currents, Eq.~\ref{eq: rho_I} and the firing rate response Eq.~\ref{eq: f-I}:
        \begin{equation*}
            \left[\nu_k^{\{m\}}\right] =  \int Dx\,\left[\nu_k^{\text{max}}\exp{\left(-\frac{(\bar{I}_k^0+\alpha_k x-\Psi_0)^2}{2\sigma_{V_k}^2}\right)}\right]^m\,,
        \end{equation*}
        where the average input current to each neuron $\bar{I}_k^0 + \alpha_k x$ is drawn from a $0$-centered Gaussian distribution of $x$ with unit variance: $D x = dx (2\pi)^{-\frac{1}{2}}\exp\left(-x^2/2\right)$. Solving the integral for $m=1$ and $m=2$, we obtain
        \begin{align}
            \bar{\nu}_k&=\nu_k^{\text{max}}\frac{\sigma_{V_k}}{\sqrt{\alpha_k^{2}+\sigma_{V_k}^{2}}}\exp\left(-\frac{(\bar{I}_k^0-\Psi_{0})^{2}}{2(\alpha_k^{2}+\sigma_{V_k}^{2})}\right)\label{eq: nu_first}\\
            q_k&=(\nu_k^{\text{max}})^2\frac{\sigma_{V_k}}{\sqrt{2\alpha_k^2+\sigma_{V_k}^2}}\exp\left(-\frac{(\bar{I}_k^0-\Psi_0)^2}{2\alpha_k^2+\sigma_{V_k}^2}\right)\label{eq: nu_second}
        \end{align}
        Standard deviations $\sigma_{V_k}$, $\alpha_k$, mean input current $\bar{I}_k^0$ and maximum firing rate $\nu_k^{\text{max}}$ contain contributions from both populations, thus forming a transcendental equation, relating first and second moments of $\nu_l$ to one another and thus defining a path towards a self-consistent solution to the model:
        
        Solving Eq.~\ref{eq: nu_first},~\ref{eq: nu_second} for $(\bar{I}_k^0-\Psi_0)^2$ both imposes a constraint on the parameter range in which self-consistent values can be found $\left(\bar{\nu}^{\star},q^{\star}\right)$ can be found (requiring real-valued input currents $I^2 \geq 0$) and defines the a relation to calculate them:
        \begin{align}
        I_k^2 &= (\bar{I}_k^0-\Psi_0)^2 \nonumber\\
    	&= \overbrace{\left(\alpha_k^2 + \sigma_{V_k}^2\right)\ln\overbrace{\left[\left(\frac{\bar{\nu}_k^*}{\nu_k^{\text{max}}}\right)^2\left(\frac{\alpha_k^2}{\sigma_{V_k}^2} + 1\right)\right]}^{<1}}^{=: \bar{I}_{\bar{\nu}}(\vec{\nu},\vec{q})} \nonumber \\
    	&\qquad = \underbrace{\left(\alpha_k^2 + \frac{1}{2}\sigma_{V_k}^2\right)\ln\underbrace{\left[\left(\frac{q_k^*}{(\nu_k^{\text{max}})^2}\right)^2\left(2\frac{\alpha_k^2}{\sigma_{V_k}^2} + 1 \right)\right]}_{<1}}_{=: \bar{I}_{q}(\vec{\nu},\vec{q})} \label{eq: selfcon}
        %&\Rightarrow \left(\frac{\bar{\nu}_k}{\nu_k^{\text{max}}}\right)^2 \left(\frac{\alpha_k^2}{\sigma_{V_k}^2} + 1\right) < 1 \; , \quad \left(\frac{q_k}{(\nu_k^{\text{max}})^2}\right)^2 \left(2\frac{\alpha_k^2}{\sigma_{V_k}^2} + 1 \right) < 1\;.
        %\quad \frac{\bar{\nu}_k}{\nu_k^{\text{max}}}\sqrt{\frac{\alpha^2}{\sigma_V^2}+1} < 1\quad,\qquad
    %       &= \left(\alpha^2 + \frac{1}{2}\sigma_{V}^2\right)\ln\left[\left(\frac{q^*}{(\nu_{\text{max}})^2}\right)^2\left(2\frac{\alpha^2}{\sigma_{V}^2} + 1 \right)\right]\nonumber\\
        %  \frac{q}{\nu_{\text{max}}^2}\sqrt{\frac{2\alpha^2}{\sigma_V^2} + 1} < 1\label{eq: inc_bound}
        \end{align}
        %with $\sigma_{V_k}^2 = \sum_{l=E,I}{\sigma_{V_{kl}}^2}$, $\alpha_k^2 = \alpha_{0,k}^2 + \sum_{l=E,I}{\alpha_{I_{kl}}^2}$ and $\nu_k^{\text{max}} = \frac{1}{2\pi} \frac{\sigma_{\dot{V}_k}}{\sigma_{V_k}}$.
        
        % We arrive at a selfconsistency equation, relating first and second moment to one another:
        % \begin{align}
        %   &\left(\tau_q (q^* + q_0) + \bar{\nu}^* \right)\ln\left[\left(\frac{\bar{\nu}^*}{\nu_{\text{max}}}\right)^2\left(\tau_q\frac{q^* + q_0}{\bar{\nu}^*} + 1\right)\right] \nonumber\\ 
        %   &\qquad = \left(\tau_q (q^*  + q_0) + \frac{\bar{\nu}^*}{2}\right)\ln\left[\left(\frac{q^*}{\nu_{\text{max}}^{2}}\right)^2\left(2\tau_{q}\frac{q^* + q_0}{\bar{\nu}^*} + 1\right)\right]\,,\label{eq: selfcon_Gauss}
        % \end{align}
        %where we used the definition of quenched ($\alpha^2$) and temporal ($\sigma_V^2$) variances as derived in the Methods section and $\tau_q = 2 (\tau_I +\tau_M)$ for brevity. 
        It is a transcendent equation that can be solved numerically exact (e.g. by a root-finding algorithm) or - in the one-population case - analytically by an approximation presented below, allowing us for the first time to present a closed form solution of the firing rate distribution in a balanced state network.

        \subsubsection{Analytical solution of the firing rate distribution}\label{sec: approx}
        
            %The last section concluded the characterization of the three dimensional phase space spanned by the three identified free parameters $\bar{\nu}$, $\alpha_0$ and $\tau_I$. To solve the selfconsistency Eq.~\ref{eq: selfcon_Gauss}, we utilized numerical methods, finding the second moment to each network averaged firing rate. However, we promised to give a closed form analytical solution, which will be discussed in the following lines.
            Self-consistent values $\left(\bar{\nu}_k^{\star},q_k^{\star}\right)$ are obtained by finding the intersection of the solutions for the input current $\bar{I}_k^0-\Psi_0$ from the evaluation of the first and second moment, Eq.~\ref{eq: selfcon}, respectively. In the case of a network composed of a single population only (thus, dropping population indices), we notice the small slope of the former, Eq.~\ref{eq: nu_first}, allowing us to approximate its solution by a constant value. Using $q=\bar{\nu}^2$ we obtain an approximate value $\bar{I}_{\bar{\nu}}^2(\bar{\nu},\bar{\nu}^2)$, not much different from the exact solution $\bar{I}_{\bar{\nu}}^2(\bar{\nu}^{\star},q^{\star})$.
            %\begin{equation*}
        	%    (\bar{I} - \Psi_0)^{2}\approx-\frac{J^{2}}{\tau_{q}}\left(\bar{\nu}+\tau_{q}\left(\bar{\nu}^{2}+q_{0}\right)\right)\ln\left[\frac{\bar{\nu}^2}{\nu_{\text{max}}^2} \left(1+\tau_{q}\left(\bar{\nu}+\frac{q_{0}}{\bar{\nu}}\right)\right)\right]
            %\end{equation*}
            %As the quadratic approximation in general underestimates the second moment $q$, the negative slope slightly overestimates the average current in the network $I_0-\Psi_0$. 
            Inserting this and the approximation of the quenched variance $\alpha_I^{2}\approx J^{2}f_{\mathcal{N}}\bar{\nu}^{2}$ (dropping $\kappa$, as it merely describes the relative sizes of different populations) into equation \ref{eq: nu_second} gives an analytically solvable expression for the second moment:
            %\begin{widetext}
            \begin{align}
        	    q &= \frac{1}{\sqrt{1+2\tau_{q}\left(\bar{\nu}+\frac{q_{0}}{\bar{\nu}}\right)}}\left(1+\tau_{q}\left(\bar{\nu}+\frac{q_{0}}{\bar{\nu}}\right)\right)^{1-\epsilon}\nu_{\text{max}}^{2\epsilon}\bar{\nu}^{2-2\epsilon} \label{eq: q_approx}\\
        	    &\text{with}\quad \epsilon=\frac{\tau_{q}\left(\bar{\nu}^{2}+q_{0}\right)}{\bar{\nu}+2\tau_{q}\left(\bar{\nu}^{2}+q_{0}\right)}\;,\qquad \tau_q = 2(\tau_I + \tau_M) \nonumber
            \end{align}
            %\end{widetext}
        %   Due to the positive slope of $(I_0-\Psi_0)(q)$, the overestimation of the current will then result in an overestimation of the second moment $q$.
            The closed form solution allows for a discussion of marginal cases, specifically in the low firing rate regime $\bar{\nu}\rightarrow 0$: A homogeneous network ($\alpha_0^2 = \frac{q_0}{J} = 0$) gives the initial approximation of a quadratic dependence $q \approx \bar{\nu}^2$. However, as soon as an arbitrary amount of inhomogeneity $q_0 \ge 0$ is added, the second moment initially grows linear with the firing rate $q\approx\frac{\nu^{\text{max}}}{\sqrt{2}}\bar{\nu}$. 
            Finally, Eq.~\ref{eq: q_approx} allows us, for the first time to our knowledge, to write down a closed form solution of a firing rate distribution of a neural network.% (Eq.~\ref{eq: rate_distribution_Gauss}).
    
    \section{Discussion}
    We presented an analytically tractable model of balanced state networks of spiking neurons in local cortical circuits. We derived closed-form equations for the firing rate distributions in one and two population networks. Neurons may be connected by synapses mediating arbitrary mixtures of slow and fast currents. In the large system limit, the shape and range of the firing rate distribution is determined by self-consistency equations for a small number of order parameters of population heterogeneity. We find that biological properties of neurons and synapses impact the emergent firing rate heterogeneity through a set of effective parameters that determine the firing rate distribution. In our model, firing rate distributions are not mathematically log-normal but can easily reproduce the form and range of skewed rate distributions observed frequently in experiments \cite{hirase_firing_2001,battaglia_firing_2005,shafi_variability_2007,hromadka_sparse_2008,oconnor_neural_2010,peyrache_spatiotemporal_2012,mizuseki_preconfigured_2013,Busche2015,shoham_how_2006,buzsaki_log-dynamic_2014}. Besides log-normal-like firing rate distributions, the model can also reproduce the monotonically decreasing rate distribution originally reported for the binary neuron model of the balanced state \cite{van_vreeswijk_chaos_1996,van_vreeswijk_chaotic_1998}. Our model should thus also enable to systematically examine the range of biological parameters consistent with realistic rate distributions and potentially predict conditions under which the generic form of rate heterogeneity breaks down.

    These advancements are based on several properties of the Gauss-Rice neuron model. Firstly, the Gauss-Rice model permits the derivation of closed-form solutions for properties such as the noisy f-I-curve or the population rate dynamic responses for arbitrary temporal correlations in the synaptic input currents with correlation functions differentiable at zero lag time \cite{tchumatchenko_correlations_2010,tchumatchenko_signatures_2010,tchumatchenko_representation_2011,tchumatchenko_spike_2011,di_bernardino_cross-correlations_2014,touzel_complete_2015}. In the current study, we used this advantage to examine networks in which cortical excitatory synapses exhibit a mix of fast and slow currents such that voltage fluctuations exhibit multiple timescales. This is neccesary to realistically model circuits in which excitatory synapses contain both AMPA and NMDA receptors, a feature of essentially every cortical network. In principle, classical LIF networks could be treated in a formally analogous fashion \cite{moreno-bote_response_2010} but only with excessive formal effort and much lower tractability. Secondly, the Gauss-Rice model with current-based synapses exhibits a Gaussian shaped noisy f-I-curve \cite{tchumatchenko_correlations_2010,tchumatchenko_signatures_2010,tchumatchenko_representation_2011,tchumatchenko_spike_2011,di_bernardino_cross-correlations_2014,touzel_complete_2015}. This feature enabled us to directly calculate the population and disorder averages in the self-consistency equations for the firing rate heterogeneity. For biological plausibility of the solutions of these self-consistency equations, only the rising part of f-I-curve should be effectively used. This condition is guaranteed in low-rate settings and can be easily validated in general. 

    Due to its great tractability, we expect that the type of circuit models introduced here will enable to advance a suite of questions and problems in modelling cortical circuits that previously were out of reach for analytical study. These include the treatment of emergent response function and tuning curve heterogeneity in sensory cortical networks, analytical heterogeneity of working memory circuits, the analysis of spatially extended, multilayer and multi-area circuits and potentially the dynamics of rate distributions in dynamic network states. For all these problems, our approach paves the way to systematically and efficiently examine the impact of quantitative cellular parameters such as membrane and synaptic time constants, AMPA to NMDA receptor ratio or cell-to-cell heterogeneity in excitability.

    \section*{Acknowledgements}
        This work was supported by the Deutsche Forschungsgemeinschaft (DFG, German Research Foundation) 436260547 in relation to NeuroNex (National Science Foundation 2015276) \& under Germany’s Excellence Strategy - EXC 2067/1- 390729940, DFG - Project-ID 317475864 - SFB 1286, DFG - Project-ID 454648639 - SFB 1528, DFG - Project-ID 273725443 - SPP 1782, DFG - Project-ID 430156276 - SPP 2205, and by the Leibniz Association (project K265/2019) (F.W.).

\bibliographystyle{IEEEtran} % We choose the "plain" reference style
%\bibliography{./bib/DarkMatter_preprint.bib} % Entries are in the refs.bib file
\bibliography{./DarkMatter_preprint.bbl} 

\begin{thebibliography}{10}
\providecommand{\url}[1]{#1}
\csname url@samestyle\endcsname
\providecommand{\newblock}{\relax}
\providecommand{\bibinfo}[2]{#2}
\providecommand{\BIBentrySTDinterwordspacing}{\spaceskip=0pt\relax}
\providecommand{\BIBentryALTinterwordstretchfactor}{4}
\providecommand{\BIBentryALTinterwordspacing}{\spaceskip=\fontdimen2\font plus
\BIBentryALTinterwordstretchfactor\fontdimen3\font minus
  \fontdimen4\font\relax}
\providecommand{\BIBforeignlanguage}[2]{{%
\expandafter\ifx\csname l@#1\endcsname\relax
\typeout{** WARNING: IEEEtran.bst: No hyphenation pattern has been}%
\typeout{** loaded for the language `#1'. Using the pattern for}%
\typeout{** the default language instead.}%
\else
\language=\csname l@#1\endcsname
\fi
#2}}
\providecommand{\BIBdecl}{\relax}
\BIBdecl

\bibitem{hirase_firing_2001}
\BIBentryALTinterwordspacing
H.~Hirase, X.~Leinekugel, A.~Czurkó, J.~Csicsvari, and G.~Buzsáki, ``Firing
  rates of hippocampal neurons are preserved during subsequent sleep episodes
  and modified by novel awake experience,'' \emph{Proceedings of the National
  Academy of Sciences}, vol.~98, no.~16, pp. 9386--9390, Jul. 2001, publisher:
  Proceedings of the National Academy of Sciences. [Online]. Available:
  \url{https://www.pnas.org/doi/10.1073/pnas.161274398}
\BIBentrySTDinterwordspacing

\bibitem{battaglia_firing_2005}
F.~P. Battaglia, G.~R. Sutherland, S.~L. Cowen, B.~L. Mc~Naughton, and K.~D.
  Harris, ``\BIBforeignlanguage{eng}{Firing rate modulation: a simple
  statistical view of memory trace reactivation},''
  \emph{\BIBforeignlanguage{eng}{Neural Networks: The Official Journal of the
  International Neural Network Society}}, vol.~18, no.~9, pp. 1280--1291, Nov.
  2005.

\bibitem{shafi_variability_2007}
M.~Shafi, Y.~Zhou, J.~Quintana, C.~Chow, J.~Fuster, and M.~Bodner,
  ``\BIBforeignlanguage{eng}{Variability in neuronal activity in primate cortex
  during working memory tasks},''
  \emph{\BIBforeignlanguage{eng}{Neuroscience}}, vol. 146, no.~3, pp.
  1082--1108, May 2007.

\bibitem{hromadka_sparse_2008}
\BIBentryALTinterwordspacing
T.~Hromádka, M.~R. DeWeese, and A.~M. Zador, ``\BIBforeignlanguage{en}{Sparse
  {Representation} of {Sounds} in the {Unanesthetized} {Auditory} {Cortex}},''
  \emph{\BIBforeignlanguage{en}{PLOS Biology}}, vol.~6, no.~1, p. e16, Jan.
  2008, publisher: Public Library of Science. [Online]. Available:
  \url{https://journals.plos.org/plosbiology/article?id=10.1371/journal.pbio.0060016}
\BIBentrySTDinterwordspacing

\bibitem{oconnor_neural_2010}
D.~H. O'Connor, S.~P. Peron, D.~Huber, and K.~Svoboda,
  ``\BIBforeignlanguage{eng}{Neural activity in barrel cortex underlying
  vibrissa-based object localization in mice},''
  \emph{\BIBforeignlanguage{eng}{Neuron}}, vol.~67, no.~6, pp. 1048--1061, Sep.
  2010.

\bibitem{peyrache_spatiotemporal_2012}
A.~Peyrache, N.~Dehghani, E.~N. Eskandar, J.~R. Madsen, W.~S. Anderson, J.~A.
  Donoghue, L.~R. Hochberg, E.~Halgren, S.~S. Cash, and A.~Destexhe,
  ``\BIBforeignlanguage{eng}{Spatiotemporal dynamics of neocortical excitation
  and inhibition during human sleep},''
  \emph{\BIBforeignlanguage{eng}{Proceedings of the National Academy of
  Sciences of the United States of America}}, vol. 109, no.~5, pp. 1731--1736,
  Jan. 2012.

\bibitem{mizuseki_preconfigured_2013}
K.~Mizuseki and G.~Buzsáki, ``\BIBforeignlanguage{eng}{Preconfigured, skewed
  distribution of firing rates in the hippocampus and entorhinal cortex},''
  \emph{\BIBforeignlanguage{eng}{Cell Reports}}, vol.~4, no.~5, pp. 1010--1021,
  Sep. 2013.

\bibitem{Busche2015}
M.~A. Busche, C.~Grienberger, A.~D. Keskin, B.~Song, U.~Neumann,
  M.~Staufenbiel, and A.~Konnerth, ``Decreased amyloid-\${\textbackslash}beta\$
  and increased neuronal hyperactivity by immunotherapy in {Alzheimer}’s
  models,'' \emph{Nature neuroscience}, vol.~18, no.~12, pp. 1725--1728, 2015.

\bibitem{shoham_how_2006}
\BIBentryALTinterwordspacing
S.~Shoham, D.~H. O’Connor, and R.~Segev, ``\BIBforeignlanguage{en}{How silent
  is the brain: is there a “dark matter” problem in neuroscience?}''
  \emph{\BIBforeignlanguage{en}{Journal of Comparative Physiology A}}, vol.
  192, no.~8, pp. 777--784, Aug. 2006. [Online]. Available:
  \url{https://doi.org/10.1007/s00359-006-0117-6}
\BIBentrySTDinterwordspacing

\bibitem{buzsaki_log-dynamic_2014}
\BIBentryALTinterwordspacing
G.~Buzsáki and K.~Mizuseki, ``\BIBforeignlanguage{en}{The log-dynamic brain:
  how skewed distributions affect network operations},''
  \emph{\BIBforeignlanguage{en}{Nature Reviews Neuroscience}}, vol.~15, no.~4,
  pp. 264--278, Apr. 2014, number: 4 Publisher: Nature Publishing Group.
  [Online]. Available: \url{https://www.nature.com/articles/nrn3687}
\BIBentrySTDinterwordspacing

\bibitem{van_vreeswijk_chaos_1996}
\BIBentryALTinterwordspacing
C.~van Vreeswijk and H.~Sompolinsky, ``Chaos in {Neuronal} {Networks} with
  {Balanced} {Excitatory} and {Inhibitory} {Activity},'' \emph{Science}, vol.
  274, no. 5293, pp. 1724--1726, Dec. 1996, publisher: American Association for
  the Advancement of Science. [Online]. Available:
  \url{https://www.science.org/doi/10.1126/science.274.5293.1724}
\BIBentrySTDinterwordspacing

\bibitem{van_vreeswijk_chaotic_1998}
------, ``\BIBforeignlanguage{eng}{Chaotic balanced state in a model of
  cortical circuits},'' \emph{\BIBforeignlanguage{eng}{Neural Computation}},
  vol.~10, no.~6, pp. 1321--1371, Aug. 1998.

\bibitem{roxin_distribution_2011}
\BIBentryALTinterwordspacing
A.~Roxin, N.~Brunel, D.~Hansel, G.~Mongillo, and C.~v. Vreeswijk,
  ``\BIBforeignlanguage{en}{On the {Distribution} of {Firing} {Rates} in
  {Networks} of {Cortical} {Neurons}},'' \emph{\BIBforeignlanguage{en}{Journal
  of Neuroscience}}, vol.~31, no.~45, pp. 16\,217--16\,226, Nov. 2011,
  publisher: Society for Neuroscience Section: Articles. [Online]. Available:
  \url{https://www.jneurosci.org/content/31/45/16217}
\BIBentrySTDinterwordspacing

\bibitem{monteforte_dynamical_2010}
\BIBentryALTinterwordspacing
M.~Monteforte and F.~Wolf, ``Dynamical {Entropy} {Production} in {Spiking}
  {Neuron} {Networks} in the {Balanced} {State},'' \emph{Physical Review
  Letters}, vol. 105, no.~26, p. 268104, Dec. 2010, publisher: American
  Physical Society. [Online]. Available:
  \url{https://link.aps.org/doi/10.1103/PhysRevLett.105.268104}
\BIBentrySTDinterwordspacing

\bibitem{monteforte_dynamic_2012}
\BIBentryALTinterwordspacing
------, ``Dynamic {Flux} {Tubes} {Form} {Reservoirs} of {Stability} in
  {Neuronal} {Circuits},'' \emph{Physical Review X}, vol.~2, no.~4, p. 041007,
  Nov. 2012, publisher: American Physical Society. [Online]. Available:
  \url{https://link.aps.org/doi/10.1103/PhysRevX.2.041007}
\BIBentrySTDinterwordspacing

\bibitem{renart_asynchronous_2010}
\BIBentryALTinterwordspacing
A.~Renart, J.~de~la Rocha, P.~Bartho, L.~Hollender, N.~Parga, A.~Reyes, and
  K.~D. Harris, ``The {Asynchronous} {State} in {Cortical} {Circuits},''
  \emph{Science}, vol. 327, no. 5965, pp. 587--590, Jan. 2010, publisher:
  American Association for the Advancement of Science. [Online]. Available:
  \url{https://www.science.org/doi/10.1126/science.1179850}
\BIBentrySTDinterwordspacing

\bibitem{amit_dynamics_1997}
\BIBentryALTinterwordspacing
D.~J. Amit and N.~Brunel, ``Dynamics of a recurrent network of spiking neurons
  before and following learning,'' \emph{Network: Computation in Neural
  Systems}, vol.~8, no.~4, pp. 373--404, Jan. 1997, publisher: Taylor \&
  Francis \_eprint: https://doi.org/10.1088/0954-898X\_8\_4\_003. [Online].
  Available: \url{https://doi.org/10.1088/0954-898X_8_4_003}
\BIBentrySTDinterwordspacing

\bibitem{brunel_dynamics_2000}
\BIBentryALTinterwordspacing
N.~Brunel, ``\BIBforeignlanguage{en}{Dynamics of {Sparsely} {Connected}
  {Networks} of {Excitatory} and {Inhibitory} {Spiking} {Neurons}},''
  \emph{\BIBforeignlanguage{en}{Journal of Computational Neuroscience}},
  vol.~8, no.~3, pp. 183--208, May 2000. [Online]. Available:
  \url{https://doi.org/10.1023/A:1008925309027}
\BIBentrySTDinterwordspacing

\bibitem{softky_highly_1993}
W.~R. Softky and C.~Koch, ``\BIBforeignlanguage{eng}{The highly irregular
  firing of cortical cells is inconsistent with temporal integration of random
  {EPSPs}},'' \emph{\BIBforeignlanguage{eng}{The Journal of Neuroscience: The
  Official Journal of the Society for Neuroscience}}, vol.~13, no.~1, pp.
  334--350, Jan. 1993.

\bibitem{otis_modulation_1992}
T.~S. Otis and I.~Mody, ``\BIBforeignlanguage{eng}{Modulation of decay kinetics
  and frequency of {GABAA} receptor-mediated spontaneous inhibitory
  postsynaptic currents in hippocampal neurons},''
  \emph{\BIBforeignlanguage{eng}{Neuroscience}}, vol.~49, no.~1, pp. 13--32,
  Jul. 1992.

\bibitem{sigel_structure_2012}
\BIBentryALTinterwordspacing
E.~Sigel and M.~E. Steinmann, ``\BIBforeignlanguage{English}{Structure,
  {Function}, and {Modulation} of {GABAA} {Receptors} *},''
  \emph{\BIBforeignlanguage{English}{Journal of Biological Chemistry}}, vol.
  287, no.~48, pp. 40\,224--40\,231, Nov. 2012, publisher: Elsevier. [Online].
  Available: \url{https://www.jbc.org/article/S0021-9258(20)62166-4/abstract}
\BIBentrySTDinterwordspacing

\bibitem{attwell_neuroenergetics_2005}
\BIBentryALTinterwordspacing
D.~Attwell and A.~Gibb, ``\BIBforeignlanguage{en}{Neuroenergetics and the
  kinetic design of excitatory synapses},''
  \emph{\BIBforeignlanguage{en}{Nature Reviews Neuroscience}}, vol.~6, no.~11,
  pp. 841--849, Nov. 2005, number: 11 Publisher: Nature Publishing Group.
  [Online]. Available: \url{https://www.nature.com/articles/nrn1784}
\BIBentrySTDinterwordspacing

\bibitem{traynelis_glutamate_2010}
S.~F. Traynelis, L.~P. Wollmuth, C.~J. McBain, F.~S. Menniti, K.~M. Vance,
  K.~K. Ogden, K.~B. Hansen, H.~Yuan, S.~J. Myers, and R.~Dingledine,
  ``\BIBforeignlanguage{eng}{Glutamate receptor ion channels: structure,
  regulation, and function},'' \emph{\BIBforeignlanguage{eng}{Pharmacological
  Reviews}}, vol.~62, no.~3, pp. 405--496, Sep. 2010.

\bibitem{compte_temporally_2003}
\BIBentryALTinterwordspacing
A.~Compte, C.~Constantinidis, J.~Tegnér, S.~Raghavachari, M.~V. Chafee, P.~S.
  Goldman-Rakic, and X.-J. Wang, ``Temporally {Irregular} {Mnemonic}
  {Persistent} {Activity} in {Prefrontal} {Neurons} of {Monkeys} {During} a
  {Delayed} {Response} {Task},'' \emph{Journal of Neurophysiology}, vol.~90,
  no.~5, pp. 3441--3454, Nov. 2003, publisher: American Physiological Society.
  [Online]. Available:
  \url{https://journals.physiology.org/doi/full/10.1152/jn.00949.2002}
\BIBentrySTDinterwordspacing

\bibitem{kerr_imaging_2005}
\BIBentryALTinterwordspacing
J.~N.~D. Kerr, D.~Greenberg, and F.~Helmchen, ``Imaging input and output of
  neocortical networks in vivo,'' \emph{Proceedings of the National Academy of
  Sciences}, vol. 102, no.~39, pp. 14\,063--14\,068, Sep. 2005, publisher:
  Proceedings of the National Academy of Sciences. [Online]. Available:
  \url{https://www.pnas.org/doi/10.1073/pnas.0506029102}
\BIBentrySTDinterwordspacing

\bibitem{dayan_theoretical_2001}
P.~Dayan and L.~F. Abbott, \emph{Theoretical neuroscience: computational and
  mathematical modeling of neural systems}, ser. Computational
  neuroscience.\hskip 1em plus 0.5em minus 0.4em\relax Cambridge, Mass:
  Massachusetts Institute of Technology Press, 2001.

\bibitem{jung_stochastic_1995}
\BIBentryALTinterwordspacing
P.~Jung, ``\BIBforeignlanguage{en}{Stochastic resonance and optimal design of
  threshold detectors},'' \emph{\BIBforeignlanguage{en}{Physics Letters A}},
  vol. 207, no.~1, pp. 93--104, Oct. 1995. [Online]. Available:
  \url{https://www.sciencedirect.com/science/article/pii/037596019500636H}
\BIBentrySTDinterwordspacing

\bibitem{naundorf_unique_2006}
B.~Naundorf, F.~Wolf, and M.~Volgushev, ``\BIBforeignlanguage{eng}{Unique
  features of action potential initiation in cortical neurons},''
  \emph{\BIBforeignlanguage{eng}{Nature}}, vol. 440, no. 7087, pp. 1060--1063,
  Apr. 2006.

\bibitem{rice_mathematical_1944}
S.~O. Rice, ``Mathematical analysis of random noise,'' \emph{The Bell System
  Technical Journal}, vol.~23, no.~3, pp. 282--332, Jul. 1944, conference Name:
  The Bell System Technical Journal.

\bibitem{tchumatchenko_signatures_2010}
\BIBentryALTinterwordspacing
T.~Tchumatchenko, T.~Geisel, M.~Volgushev, and F.~Wolf, ``Signatures of
  {Synchrony} in {Pairwise} {Count} {Correlations},'' \emph{Frontiers in
  Computational Neuroscience}, vol.~4, p.~1, Apr. 2010. [Online]. Available:
  \url{https://www.ncbi.nlm.nih.gov/pmc/articles/PMC2857958/}
\BIBentrySTDinterwordspacing

\bibitem{tchumatchenko_correlations_2010}
\BIBentryALTinterwordspacing
T.~Tchumatchenko, A.~Malyshev, T.~Geisel, M.~Volgushev, and F.~Wolf,
  ``Correlations and {Synchrony} in {Threshold} {Neuron} {Models},''
  \emph{Physical Review Letters}, vol. 104, no.~5, p. 058102, Feb. 2010,
  publisher: American Physical Society. [Online]. Available:
  \url{https://link.aps.org/doi/10.1103/PhysRevLett.104.058102}
\BIBentrySTDinterwordspacing

\bibitem{tchumatchenko_representation_2011}
\BIBentryALTinterwordspacing
T.~Tchumatchenko and F.~Wolf, ``\BIBforeignlanguage{en}{Representation of
  {Dynamical} {Stimuli} in {Populations} of {Threshold} {Neurons}},''
  \emph{\BIBforeignlanguage{en}{PLOS Computational Biology}}, vol.~7, no.~10,
  p. e1002239, Oct. 2011, publisher: Public Library of Science. [Online].
  Available:
  \url{https://journals.plos.org/ploscompbiol/article?id=10.1371/journal.pcbi.1002239}
\BIBentrySTDinterwordspacing

\bibitem{tchumatchenko_spike_2011}
\BIBentryALTinterwordspacing
T.~Tchumatchenko, T.~Geisel, M.~Volgushev, and F.~Wolf, ``Spike {Correlations}
  – {What} {Can} {They} {Tell} {About} {Synchrony}?'' \emph{Frontiers in
  Neuroscience}, vol.~5, 2011. [Online]. Available:
  \url{https://www.frontiersin.org/articles/10.3389/fnins.2011.00068}
\BIBentrySTDinterwordspacing

\bibitem{di_bernardino_cross-correlations_2014}
\BIBentryALTinterwordspacing
E.~Di~Bernardino, J.~León, and T.~Tchumatchenko, ``Cross-{Correlations} and
  {Joint} {Gaussianity} in {Multivariate} {Level} {Crossing} {Models},''
  \emph{The Journal of Mathematical Neuroscience}, vol.~4, no.~1, p.~22, Apr.
  2014. [Online]. Available: \url{https://doi.org/10.1186/2190-8567-4-22}
\BIBentrySTDinterwordspacing

\bibitem{touzel_complete_2015}
\BIBentryALTinterwordspacing
M.~P. Touzel and F.~Wolf, ``\BIBforeignlanguage{en}{Complete {Firing}-{Rate}
  {Response} of {Neurons} with {Complex} {Intrinsic} {Dynamics}},''
  \emph{\BIBforeignlanguage{en}{PLOS Computational Biology}}, vol.~11, no.~12,
  p. e1004636, Dec. 2015, publisher: Public Library of Science. [Online].
  Available:
  \url{https://journals.plos.org/ploscompbiol/article?id=10.1371/journal.pcbi.1004636}
\BIBentrySTDinterwordspacing

\bibitem{moreno-bote_response_2010}
R.~Moreno-Bote and N.~Parga, ``\BIBforeignlanguage{eng}{Response of
  integrate-and-fire neurons to noisy inputs filtered by synapses with
  arbitrary timescales: firing rate and correlations},''
  \emph{\BIBforeignlanguage{eng}{Neural Computation}}, vol.~22, no.~6, pp.
  1528--1572, Jun. 2010.

\end{thebibliography}

\clearpage

\appendix

\section{Appendices}

    \subsection{Variance of the compound spike train}\label{sec: appendix_compound_spike_train_variance}
    Due to independence of the connectivity matrix $A_{kl}^{ij}$, the solutions to the compound spike train variance for different populations decouple and can be obtained as a superposition of the solutions to the distinct populations. 
    
    The second moment for one population yields: 
    \begin{align*}
        [\Omega_{i}^{2}]_i&=\left[\left(\sum_{j=1}^{N}A_{ij}\nu_{j}\right)\left(\sum_{k=1}^{N}A_{ik}\nu_{k}\right)\right]_{i}\\
        &=\sum_{j=1}^{N}\sum_{k=1}^{N}\left[A_{ij}A_{ik}\right]_{i}\nu_{j}\nu_{k}
    \end{align*}
    This can be split into a diagonal ($j=k$) and an off-diagonal ($j\neq k$) term:
    \begin{align*}
        [\Omega_{i}^{2}]_i&=\sum_{j=1}^{N}[A_{ij}]_{i}\nu_{j}^2 + \sum_{j=1}^{N}\sum_{k\neq j}[A_{ij}]_{i}[A_{ik}]_{i}\nu_{j}\nu_{k}\\
        &= K\left[\nu_j^2\right]_j + \sum_{j=1}^{N}[A_{ij}]_i\nu_{j}\sum_{k=1}^{N}[A_{ik}]_{i}\nu_{k}-\sum_{j=1}^{N}[A_{ij}]_i^2\nu_{j}^2\\
        &= K\left[\nu_j^2\right]_j + K^2 \bar{\nu}^2 - \frac{1}{N}[\left(K^{\text{out}}_j\right)^2]_j[\nu_{j}]_j^2
    \end{align*}
    As the out-degree follows a poisson distribution, the second moment is $K^{2}+K$, which vanishes against $N$ and the variance is obtained as:
    \begin{equation*}
        \text{Var}\left(\Omega\right) = [\Omega_{i}^{2}]_i-[\Omega_{i}]_i^{2} = K [\nu_j^{2}]_j + K^2 \bar{\nu}^2 - K^2\bar{\nu}^2 = Kq \,,
    \end{equation*}
    where $q$ denotes the second moment of the firing rate distribution. The solution to multiple populations $l$ is obtained by substituting $K\rightarrow \kappa_l K$ for each population, resulting in:
    \begin{equation*}
        \text{Var}\left(\Omega\right) = K \sum_l \kappa_l q_l \,.
    \end{equation*}

    \subsection{Input current}\label{sec: appendix_input_current}
    \subsubsection{Averaged input current}
    Similar to App.~\ref{sec: appendix_compound_spike_train_variance}, we can obtain the average input current to neuron $i$ as the sum of average input currents from either population, following from the statistics of the compound spike train, Eq.~\ref{eq: methods_compound_spike_train}.

    The probability of a specific spike train realization from a single population is
    \begin{equation*}
	    P\left(\{ t^{(p)}\}\mid\Omega(t)\right)=\frac{e^{-\int_{0}^{T}\D t \Omega(t)}}{N_{\text{AP}}!}\prod_{p=1}^{N_{\text{AP}}}\Omega(t^{(p)})\,,
    \end{equation*}
    with the factor $1/N_{\text{AP}}!$ accounting for the unordered nature of such randomized sequences. As the PSCs are assumed to be uncorrelated, the integration over spike times $t^{(p)}$ is interchangeable. The average current within time interval $t\in [0,T]$ is
      \begin{widetext}
          \begin{align*}
    	    \langle I_{i}(t)\rangle &= \sum_{N_{\text{AP}}=0}^{\infty}\int_{0}^{T}\dots\int_{0}^{T}\left(\prod_{p=1}^{N_{\text{AP}}}\D t ^{(p)}\right)I\left(t,\{ t^{(p)}\}\right)P\left(\{ t^{(p)}\mid\Omega_i(t^{(p)}\}\right) \\
    	    &= \sum_{N_{\text{AP}}=0}^{\infty}\int_{0}^{T}\dots\int_{0}^{T}\left(\prod_{p=1}^{N_{\text{AP}}}\D t ^{(p)}\right)\left(J\sum_{p'=1}^{N_{\text{AP}}}f(t-t^{(p')})\right)\frac{e^{-\int_{0}^{T}\D s \Omega_i(s)}}{N_{\text{AP}}!}\prod_{p''=1}^{N_{\text{AP}}}\Omega_{i}(t^{(p'')}) \\
    	    &= J e^{-\int_{0}^{T}\D s \Omega_i(s)}\sum_{N_{\text{AP}}=1}^{\infty}\frac{1}{(N_{\text{AP}} -1)!}\left(\int_{0}^{T}\D s \Omega_{i}(s)\right)^{N_{\text{AP}}-1}\int_{0}^{T}\D t ^{(1)}f(t-t^{(1)})\Omega_{i}(t^{(1)})\,,
          \end{align*}
      \end{widetext}
      where $\langle \cdot \rangle$ denotes time average. The last line uses that all $N_{\text{AP}}$ integrals over one PSC yield the same result, as long as the kernel $f(t-t^{(p)})$ is within the interval $t\in[0,T]$. As the PSCs are equal, we can chose $t^{(1)}$ to be representative of all $N_{\text{AP}}$ events. The sum - neglecting the first, vanishing term of $N_{\text{AP}} = 0$ - is the definition of the exponential series, canceling with the previous term and the equation then reads
      \begin{equation}
	    \langle I_{i}(t)\rangle=\frac{J}{\sqrt{K}} \int_{0}^{T}\D t ^{(1)}f(t-t^{(1)})\Omega_{i}(t^{(1)}) = \frac{J}{\sqrt{K}} f_{\mathcal{N}} \Omega_i\, ,
      \end{equation}
      where the last equation holds in the stationary case ($\Omega_{i}(t^{(1)}) = \Omega_{i}$), with the $f_{\mathcal{N}}$ the integral over the kernel.% (=1 for $\mathcal{N}=\tau_{I_{l_m}}$, = $\tau_{I_{l_m}}$ for $\mathcal{N}=1$).
    
    \subsubsection{Input current covariance}\label{sec: appendix_input_current_covariance}
    The statistics of input current temporal fluctuations can be captured by the input current covariance function:
    \begin{align*}
        C_{I_{i}}(t,t') &:= \langle\delta I^{i}(t)\delta I^{i}(t')\rangle\\
	    &= 2\underbrace{\langle \delta I_E^{i}(t)\delta I_I^{i}(t) \rangle}_{\text{uncorrelated} \rightarrow 0} + \sum_{l=E,I} \langle \delta I_l^{i}(t)\delta I_l^{i}(t')\rangle\,.
    \end{align*}
    We can thus evaluate the covariance for each population separately (dropping population indices):
    \begin{align}
	    C_{I_{i}}(t,t') &:= \langle\delta I_{i}(t)\delta I_{i}(t')\rangle \nonumber\\
	    &=\langle I_{i}(t)I_{i}(t')\rangle-\langle I_{i}(t)\rangle\langle I_{i}(t')\rangle \label{eq: I_corr_def}
    \end{align}
    Here $\delta I_i(t)$ is the deviation of \(I_i(t)\) from the time average \(\langle I_{i}(t)\rangle\). The second term is already known from Eq.~\ref{eq: I_tmp_avg} and the first one can be obtained as:
%       For better readability we drop the subscript $i$ below each $I$ and $\Omega$. We compute the ensemble average\begin{eqnarray}
    %\begin{widetext}
      \begin{align*}
        \left\langle I(t)I(t')\right\rangle = \left\langle J\sum_{p'=1}^{N_{\text{AP}}}f\left(t-t^{(p')}\right)J\sum_{p''=1}^{N_{\text{AP}}}f\left(t'-t^{(p'')}\right)\right\rangle \\%=J^{2}\left\langle \sum_{p',p''=1}^{N_{\text{AP}}}f\left(t-t^{(p')}\right)f\left(t'-t^{(p'')}\right)\right\rangle \nonumber \\
        = \frac{J^{2}}{K}\sum_{N_{\text{AP}}=0}^{\infty}\frac{e^{-\int_{0}^{T}\D s \Omega_i(s)}}{N_{\text{AP}}!}\int_{0}^{T}\dots\int_{0}^{T}\left(\prod_{p=1}^{N_{\text{AP}}}\D t ^{(p)}\Omega_i\left(t^{(p)}\right)\right) \\ \qquad \times \sum_{p'=1}^{N_{\text{AP}}}\sum_{p''=1}^{N_{\text{AP}}}f\left(t-t^{(p')}\right)f\left(t'-t^{(p'')}\right)%\label{eq:I_correlation}
      \end{align*}
      %\end{widetext}
      Again, the term $N_{\text{AP}}=0$ gives no contribution. We continue with a distinction between diagonal ($p'=p''$) and off-diagonal ($p'\neq p''$) terms, evaluating the former first:
      % and only the timeshifted PSCs from the same spike time (terms with $p'= p''$) give a contribution, as they are evaluated along the same dimension of the hypercube:
      %\begin{widetext}
      \begin{align*}
	\left\langle I(t)I(t')\right\rangle_{p'=p''} %&= J^{2}\sum_{N_{\text{AP}}=1}^{\infty}\frac{e^{-\int_{0}^{T}\D s \Omega(s)}}{N_{\text{AP}}!}\int_{0}^{T}\left(\prod_{p=1}^{N_{\text{AP}}}\D t ^{(p)}\Omega\left(t^{(p)}\right)\right)\sum_{p'=1}^{N_{\text{AP}}}f\left(t-t^{(p')}\right)f\left(t'-t^{(p')}\right)\\
	= \frac{J^{2}}{K}\sum_{N_{\text{AP}}=1}^{\infty}\frac{e^{-\int_{0}^{T}\D s \Omega_i(s)}}{(N_{\text{AP}}-1)!}\left(\int_{0}^{T}\D s \Omega_i(s)\right)^{N_{\text{AP}}-1} \nonumber\\
	\qquad \times \int_{0}^{T}\D t ^{(1)}f\left(t-t^{(1)}\right)f\left(t'-t^{(1)}\right)\Omega_i\left(t^{(1)}\right)\nonumber\\
	\quad = \frac{J^{2}}{K}\int_{0}^{T}\D t ^{(1)}f\left(t-t^{(1)}\right)f\left(t'-t^{(1)}\right)\Omega_i(t^{(1)})\,, \label{eq: I_corr_diag}
      \end{align*}
      %\end{widetext}
      again using the identity of all $N_{\text{AP}}$ terms and the series representation of the exponential function. 
      
      Calculating the crosscorrelation of PSCs (off-diagonal terms) involves spike trains with two or more APs. Therefor the double sum contains $N_{\text{AP}}(N_{\text{AP}}-1)$ terms which, due to the independence of spike times factorize:
      \begin{widetext}
          \begin{align*}
          \left\langle I(t)I(t')\right\rangle_{p'\neq p''} %&= J^{2}\sum_{N_{\text{AP}}=2}^{\infty}\frac{e^{-\int_{0}^{T}\D s \Omega(s)}}{N_{\text{AP}}!}\int_{0}^{T}\left(\prod_{p=1}^{N_{\text{AP}}}\D t ^{(p)}\Omega\left(t^{(p)}\right)\right)\sum_{p'\neq p''}f\left(t-t^{(p')}\right)f\left(t'-t^{(p'')}\right)\\
          & = \frac{J^{2}}{K} \sum_{N_{\text{AP}}=2}^{\infty}\frac{e^{-\int_{0}^{T}\D s \Omega_i(s)}}{(N_{\text{AP}}-2)!}\left(\int_{0}^{T}\D s \Omega_i(s)\right)^{N_{\text{AP}}-2} \int_{0}^{T}\D t ^{(1)}f\left(t-t^{(1)}\right)\Omega_i\left(t^{(1)}\right) \int_{0}^{T}\D t ^{(2)}f\left(t'-t^{(2)}\right)\Omega_i\left(t^{(2)}\right)\\
          & = \frac{J^{2}}{K}\int_{0}^{T}\D t ^{(1)}f(t-t^{(1)})\Omega_i(t^{(1)})\int_{0}^{T}\D t ^{(2)}f(t'-t^{(2)})\Omega_i(t^{(2)}) = \langle I(t)\rangle\langle I(t')\rangle %\label{eq: I_corr_offdiag}
          \end{align*}
      \end{widetext}
      The off-diagonal terms thus cancel the latter term in Eq.~\ref{eq: I_corr_def} and the temporal covariance function of the input current of a single population evaluates to:
%       We finally obtain the temporal correlation of the input current in the stationary case from Eq.~\ref{eq: I_tmp_avg},~\ref{eq: I_corr_diag} and \ref{eq: I_corr_offdiag}:
      \begin{align}
	C_{I}(t,t')=\langle\delta I(t)\delta I(t')\rangle=\frac{J^{2}}{K}\Omega_{i}\int_{-\infty}^{\infty}\D s f(t-s)f(t'-s)
      \end{align}

    \subsection{Covariance of the membrane potential}\label{sec: appendix_potential_covariance}
        
        The membrane potential dynamics are driven by the input current dynamics, derived in App.~\ref{sec: appendix_input_current}, which are processed by the LIF-specific membrane potential response function $G(\Delta t)=\frac{1}{\tau_M}e^{-\frac{\Delta t}{\tau_M}}$.
        
        Calculating the crosscorrelation of voltage and current and the autocorrelation of $G(\Delta t)$ beforehand allows the subsequent evaluation of the membrane potential autocorrelation:
        \begin{align*}
            \langle\delta V(t)\delta I(s')\rangle
            &=\langle\int_{-\infty}^{+\infty}\D s G(t-s)\delta I(s)\delta I(s')\rangle \nonumber\\
            &=\int_{-\infty}^{+\infty}\D s G(t-s)C_{I}(s'-s)\\%\label{eq: V_I_corr}\\
            G^{(2)}(t)&:=\int_{-\infty}^{+\infty}\D s G(s)G(s-t)=\frac{1}{2\tau_M}\exp\left(-\frac{|t|}{\tau_M}\right)\\%\label{eq: G_G_corr}
        \end{align*}
        $C_V(t)$ now calculates as:
        \begin{align*}
        	C_{V}(t) &= \langle\delta V(t)\delta V(0)\rangle = \langle\delta V(t)\int_{-\infty}^{+\infty}\D s 'G(s')\delta I(-s')\rangle=\\
        	&= \int_{-\infty}^{+\infty}\D s 'G(s')\langle\delta V(t)\delta I(-s')\rangle\\
        	&=\int_{-\infty}^{+\infty}\D s 'G(s')\int_{-\infty}^{+\infty}\D s G(t-s)C_{I}(-s'-s)\\
        	&= \int_{-\infty}^{+\infty}\D s C_{I}(s)\int_{-\infty}^{+\infty}\D s 'G(s')G(s'-s+t)\\
        	&=\int_{-\infty}^{+\infty}\D s C_{I}(s)G^{(2)}(s-t)\\
        	&= \frac{\sigma_{I}^{2}}{2\tau_M}\int_{-\infty}^{+\infty}\D s \exp\left(-\frac{|s|}{\tau_{I_{l}}}\right)\exp\left(-\frac{|s-t|}{\tau_M}\right)\,,
        \end{align*}
        using a mono-synaptic kernel, Eq.~\ref{eq: kernel_single}.
      %Symmetry of the autocorrelation function allows us to consider the case of $t>0$ only for evaluating this expression and thus obtain the general solution for $|t|$. 
        The integral splits up into three regions: $s<0$, $0<s<t$ and $s>t$ and we obtain distinct solutions for the covariance contribution of population $l$ for the case of $\tau_{I_{l}} \neq \tau_M$:
        \begin{align*}
    	    %\tau_I \neq \tau_M:\quad 
    	    C_{V_l}(\Delta t) &= \frac{\sigma_{I}^{2}\tau_{I_l}}{\tau_{I_l}^{2}-\tau_M^{2}}\left[\tau_{I_l}\exp\left(-\frac{|\Delta t|}{\tau_{I_l}}\right)-\tau_M\exp\left(-\frac{|\Delta t|}{\tau_M}\right)\right]\\
        \intertext{and in the case of $\tau_{I_{l}} = \tau_M$:}
            %\tau_{I_l} = \tau_M:\quad 
            C_{V_l}(\Delta t) &= \frac{\sigma_{I}^{2}}{2}\left(1+\frac{\left|\Delta t\right|}{\tau_M}\right)\exp\left(-\frac{\left|\Delta t\right|}{\tau_M}\right)\nonumber
        \end{align*}

%\clearpage

\end{document}